\documentclass[]{article}
\usepackage{multicol}
\usepackage{apj}
\def\lesssim{\mathrel{\hbox{\rlap{\hbox{\lower4pt\hbox{$\sim$}}}\hbox{$<$}}}}
\def\gtrsim{\mathrel{\hbox{\rlap{\hbox{\lower4pt\hbox{$\sim$}}}\hbox{$>$}}}}
\newcommand{\mincir}{\raise -2.truept\hbox{\rlap{\hbox{$\sim$}}\raise5.truept
\hbox{$<$}\ }}
\newcommand{\magcir}{\raise -2.truept\hbox{\rlap{\hbox{$\sim$}}\raise5.truept
\hbox{$>$}\ }}
\newcommand{\siml}{\raise -2.truept\hbox{\rlap{\hbox{$\sim$}}\raise5.truept
\hbox{$<$}\ }}
\newcommand{\simg}{\raise -2.truept\hbox{\rlap{\hbox{$\sim$}}\raise5.truept
\hbox{$>$}\ }}
\newcommand{\be}{\begin{equation}}
\newcommand{\ee}{\end{equation}}
\newcommand{\ba}{\begin{eqnarray}}
\newcommand{\ea}{\end{eqnarray}}
\newcommand {\h} {$h^{-1}$ Mpc $ \;$}

\newcommand {\hh} {$h^{-1}$ Mpc}
\newcommand {\ks} {km~s$^{-1} \;$}
\newcommand {\kss} {km~s$^{-1}$}
\newcommand {\msun} {$h^{-1} \  M_{\odot} \;$}

\newcommand {\ml} {$h \, M_{\odot}/L_{\odot} \;$}
\newcommand {\mll} {$h \, M_{\odot}/L_{\odot}$}

\begin{document}

\vspace{15mm}
\begin{center}
\uppercase{Observational
Mass--to--Light Ratio \\ of Galaxy Systems:\\
from Poor Groups to Rich Clusters} 
\vspace*{1.5ex}
{\sc Marisa Girardi$^{1}$, Patrizia Manzato$^{1}$, Marino Mezzetti$^{1}$,
Giuliano Giuricin$^{2}$, and F\"usun Limboz$^{3}$}\\
\vspace*{1.ex}
{\small
$^1$ Dipartimento di Astronomia, Universit\`{a} 
degli Studi di Trieste, Via Tiepolo 11, I-34131 Trieste, Italy;
E-mail: girardi, manzato, mezzetti@ts.astro.it\\
$^2$ Deceased. Formerly at Dipartimento di Astronomia, Universit\`{a} 
degli Studi di Trieste\\
$^3$ Instanbul University, Science Faculty, Astronomy 
and Space Science Dept., 34452 Beyazit, Instanbul, Turkey}\\
\end{center}  
\vspace*{-6pt}

\begin{abstract}
We study the mass--to--light ratio of galaxy systems from poor groups
to rich clusters, and present for the first time a large database for
useful comparisons with theoretical predictions.  We extend a previous
work, where $B_j$ band luminosities and optical virial masses were
analyzed for a sample of 89 clusters. Here we also consider a sample
of 52 more clusters, 36 poor clusters, 7 rich groups, and two
catalogs, of $\sim 500$ groups each, recently identified in the Nearby
Optical Galaxy sample by using two different algorithms.  We obtain
the blue luminosity and virial mass for all systems considered. We
devote a large effort to establishing the homogeneity of the resulting
values, as well as to considering comparable physical regions, i.e.
those included within the virial radius.  By analyzing a fiducial,
combined sample of 294 systems we find that the mass increases faster
than the luminosity: the linear fit gives $M\propto L_{B}^{1.34 \pm
0.03}$, with a tendency for a steeper increase in the low--mass range.
In agreement with the previous work, our present results are superior
owing to the much higher statistical significance and the wider
dynamical range covered ($\sim 10^{12}$--$10^{15}~M_{\odot}$).  We
present a comparison between our results and the theoretical
predictions on the relation between $M/L_B$ and halo mass, obtained by
combining cosmological numerical simulations and semianalytic modeling
of galaxy formation.

%\end{abstract}
%\keywords{ galaxies: clusters: general -- galaxies: fundamental 
%parameters -- cosmology: observations}

\vspace*{6pt}
\noindent
{\em Subject headings: }
galaxies: clusters: general -- galaxies: fundamental 
parameters -- cosmology: observations.
\end{abstract}

\begin{multicols}{2}

\section{INTRODUCTION}

Since the work by Zwicky (1933), it is well known that the luminous
matter associated with galaxies in clusters provides only a small part
of the total cluster mass. The relative contribution of the dark
matter component is usually specified in terms of the mass--to--light
ratio, $M/L$, the total amount of mass relative to the
total light within a given scale.  

Pioneering analyses showed that $M/L$ increases from the bright
luminous parts of galaxies to cluster scales (Blumenthal et al. 1984).
Indeed, models of biased galaxy formation, where galaxies formed only
in the highest peaks in the initial fluctuation spectrum, naturally
predict an increase of $M/L$ with system mass (e.g., Bardeen et
al. 1986; Davis et al. 1985).  Several mechanisms whereby the
efficiency of galaxy formation is biased towards very high density
peaks are possible (e.g., Rees 1985).  However, only recently has the
combination of cosmological N--body simulations and semianalytic
modeling of galaxy formation allowed realistic predictions about the
$M/L$ of galaxy systems (Kauffmann et al. 1999; Bahcall et al. 2000;
Benson et al. 2000; Somerville et al. 2001). Although differing in
details, it has been generally found that $M/L$ increases with mass
halo from very poor to rich systems, possibly with a flattening on
large scales.

As for the observational point of view, the estimate of $M/L$ in
galaxy systems is not an easy task.  Both mass and luminosity
estimates are fraught with several uncertainties.  The uncertainties
in the luminosity determination are related to corrections for
calibration of the photometry (when using inhomogeneous photometric
data), background galaxy contamination, and the need to extrapolate
the sum of measured luminosities of galaxy members to include faint
galaxies and the outer parts of the systems, beyond the region studied
(see, e.g., Oemler 1974).

Also the estimate of masses is not an easy task, in spite of the
various methods which have been applied (e.g., Narayan \& Bartelmann
1996; Schindler 1996; Mellier 1999; Biviano 2001).  Masses of galaxy
systems are inferred from either $X$--ray or optical data, under the
general hypothesis of dynamical equilibrium.  Estimates based on
gravitational lensing do not require assumptions about the dynamical
status of the system, but a good knowledge of the geometry of the
potential well is necessary.  Claims for a discrepancy (by a factor of
2--3) between cluster masses obtained with different methods cast
doubts about the general reliability of mass estimates (e.g., Wu \&
Fang 1997).  However, recent analyses have shown that, if we avoid
cases of bimodal clusters, mass estimates concerning large cluster
areas are in general agreement (Allen 1997; Girardi et al. 1998b,
hereafter G98; Lewis et al. 1999).

Large collections of observational data concerning galaxies, groups,
and clusters suggest that all systems have a constant ratio of
$M/L_B\sim 200$--$300$ \ml for scales larger than galaxies, so that
the total mass of galaxy systems could be roughly accounted for by the
total mass of their member galaxies, possibly plus the mass of the hot
intracluster gas (Rubin 1993; Bahcall, Lubin, \& Dorman 1995).
Homogeneous samples, where both masses and luminosities are computed
in a consistent way, would be more reliable.  Unfortunately, the above
observational difficulties prevented us from building a large $M/L$
data base spanning a wide dynamical range.  Based on homogeneous
optical data, the pioneering work by Dressler (1978) showed no
evidence of correlation of $M/L$ values with richness for 12
clusters. More recently, David, Jones, \& Forman (1995), who used
homogeneous X--ray mass estimates and luminosities from different
sources in the literature, showed that $M/L_V$ of seven groups and
clusters of galaxies are comparable. Also $M/L_r$ values for the
sample of 15 clusters of the Canadian Network for Observational
Cosmology (CNOC, Calberg et al. 1996), where masses come from optical
virial estimates, are consistent with an universal underlying value.

A slight increase of $M/L$ with mass system was suggested by indirect
analyses of the cluster fundamental plane, i.e., the study of the
relations between cluster size, internal velocity dispersion, and
luminosity (but see Fritsch \& Buchert 1999). In fact, assuming the
virialization state and internal structure of all clusters to be
identical, one can derive the behavior of $M/L$. Working with a
homogeneous photometric sample of 12 clusters Schaeffer et al. (1993)
found that $M/L_V\propto L_V^{0.3}$.  Similarly, using homogeneous
results for 29 clusters of the ESO Nearby Abell Cluster Survey (ENACS;
Katgert et al. 1998), Adami et al. (1998a) showed that
$M/L_{B_j}\propto \sigma_v$, where $\sigma_v$ is the line--of--sight
(l.o.s.) velocity dispersion of member galaxies. This correlation was
also directly verified by Adami et al. (1998b) in a following work by
computing the projected virial masses.

Recently, Girardi et al. (2000, hereafter G00) faced the question from
a direct point of view with a significant increment in the data--base
statistics.  They analyzed 89 clusters, all with homogeneous optical
virial mass estimates by G98 and homogeneous luminosity estimates
derived from the COSMOS catalog (Yentis et al. 1992). Moreover, the
available data allowed the authors to compute mass and luminosity
within the virial radius in order to analyze physically comparable
regions in poor and rich clusters.  Their main result is that the mass
has a slight, but significant tendency to increase faster than the
luminosity: $M\propto L_{B_j}^{\mbox{\rm 1.2--1.3}}$.  Owing to the
large uncertainties generally involved, it is really not surprising
that such a slight effect could not be detected by previous analyses
based on small statistics and/or a small dynamical range and/or
inhomogeneous samples.

Recent support for G00 results came from a study of $\sim$ 200 galaxy
groups, identified within the field galaxy redshift survey CNOC2 by
Carlberg et al. (2001a), showing evidence that $M/L$ increases with
increasing $\sigma_v$. Moreover, $M/L$ of CNOC2 groups proves to be
smaller than that of CNOC clusters (Hoekstra, Yee, \& Gladders
2001). However, the question is still open.  E.g., Hradecky et
al. (2000), who computed homogeneous X--ray mass and optical
luminosity for eight galaxy systems, claimed that $M/L_V$ is roughly
independent of system mass.

New insights on the behavior of $M/L$ for galaxy systems of different
mass would be particularly useful in view of the theoretical
predictions recently coming from cosmological N--body simulations
combined with semianalytic modeling of galaxy formation.  To draw more
definitive conclusions about this topic, we extend the work of
G00 by increasing the statistics of the data base and doubling the
dynamical range, from $\sim 5\times 10^{13}$--$10^{15}$ \msun to $\sim
10^{12}$--$10^{15}$ \msun.  To this purpose, we consider both
clusters analyzed by G98, the poor clusters by Ledlow et al. (1996;
hereafter L96), the rich groups by Zabludoff \& Mulchaey (1998a;
hereafter ZM98), and the groups identified in the NOG sample (Nearby
Optical Galaxy, Giuricin et al. 2000).

The paper is organized as follows.  We describe the data samples in
\S~2.  We compute the main observational quantities, i.e. virial
masses and optical luminosities, for all galaxy systems in \S~3. We
devote \S~4 to the analysis of the relation between mass and
luminosity and to the mass--to--light ratio.  We discuss our results
in \S~5, while in \S~6 we give a brief summary of our main results and
draw our conclusions.

Unless otherwise stated, we give errors at the 68\% confidence
level (hereafter c.l.)

A Hubble constant of 100 $h$ \ks $Mpc^{-1}$ is used throughout.

\section{DATA SAMPLES}

Table~1 briefly summarizes the samples of galaxy systems in this
work. We list: the sample name with the corresponding number of
systems, $N_S$, and references for the catalog [Cols.~(1), (2), and
(3), respectively]; the subsample name with the corresponding number
of systems, $N_{SS}$ [Cols.~(4) and (5), respectively]; the references
for galaxy redshift and coordinates used for mass determination
[Col.~(6)]; the references for galaxy magnitudes used for luminosity
determination [Col.~(7)]; and a brief description of the sample
[Col.~(8)].  Detailed comments are given below.

\end{multicols}
%%TAB1%\vspace{-6mm}
\hspace{-13mm}
\begin{minipage}{20cm}
\renewcommand{\arraystretch}{1.2}
\renewcommand{\tabcolsep}{1.2mm}
\begin{center}%\vspace{-3mm}
TABLE 1\\%\vspace{2mm}
{\sc Samples of Galaxy Systems \\}
\footnotesize 

%%%%%%%%%%%%%%%%%%%%%%%%%%%%%%%%%%%%%%%%%%%%%%%%%%%%%%%%%%%%%%%%%%%%%%%
%
%                        TABLE 1
%
%%%%%%%%%%%%%%%%%%%%%%%%%%%%%%%%%%%%%%%%%%%%%%%%%%%%%%%%%%%%%%%%%%%%%%%
  
\begin{tabular}{lrccrccl}
\hline \hline
\multicolumn{1}{c}{Name} 
&\multicolumn{1}{c}{$N_S$}
&\multicolumn{1}{c}{Refs.}  
&\multicolumn{1}{c}{Name$_{SS}$} 
&\multicolumn{1}{c}{$N_{SS}$}
&\multicolumn{1}{c}{Cord./z Refs.\tablenotemark{a}}  
&\multicolumn{1}{c}{Phot. Refs.}
&\multicolumn{1}{c}{Description} 
\\ 
\multicolumn{1}{c}{(1)} 
&\multicolumn{1}{c}{(2)} 
&\multicolumn{1}{c}{(3)}
&\multicolumn{1}{c}{(4)} 
&\multicolumn{1}{c}{(5)} 
&\multicolumn{1}{c}{(6)} 
&\multicolumn{1}{c}{(7)}
&\multicolumn{1}{c}{(8)} 
\\
\hline
CL &119& G98 &C-CL& 89& G98& COSMOS& clusters (mainly ACO)\\ 
CL &119& G98 &A-CL& 52& G98& APS & clusters (mainly ACO)\\ 
PS & 43& L96,ZM98&C-PS& 8& NED,ZM98 & COSMOS& poor clusters, rich groups \\ 
PS & 43& L96,ZM98&A-PS& 40& NED,ZM98 & APS &poor clusters, rich groups \\ 
HG &475& NOG & & & NOG & NOG &loose groups \\ 
PG &513& NOG & & & NOG & NOG &loose groups \\ 
\hline
\end{tabular}

\end{center}
\vspace{-2mm}
{\footnotesize\parindent=3mm 
$^a$~References for galaxy data are 
those which summarize information, e.g. the original redshift data for
clusters come from ENACS and other literature: the reader can
find them in G98.}

\end{minipage}
\begin{multicols}{2}

\subsection{Cluster Sample (CL)}

The sample of nearby clusters ($z\le 0.15$) of G98 is an extension of
that of Fadda et al. (1996) and collects clusters having at least 30
galaxies with available redshifts in the field, in order  to allow
homogeneous and robust estimates of internal velocity dispersion and
cluster mass. G00 have already analyzed a subsample of 89 clusters
for which galaxy magnitudes are available in the COSMOS catalog
(hereafter C-CL sample).

Here we select another 52 clusters (hereafter A-CL sample) for which
galaxy magnitudes are available in the Revised APS catalog of POSSI
(Pennington et al. 1993).  In particular, we avoid the G98 clusters
which show two peaks either in the velocity or in the projected galaxy
distribution, as well as clusters with uncertain dynamics (cf. \S~2 of
G98). From the G98 analysis we take for each cluster: the
l.o.s. velocity dispersion $\sigma_v$, the cluster center, the virial
radius $R_{vir}$ (there called virialization radius), and the
(corrected) virial mass $M$ computed within $R_{vir}$.

Among A-CLs, 22 systems are in common with C-CLs and will
be used to homogenize the photometric data.

\subsection{Sample of Poor Systems   (PS)}

The 71 poor galaxy clusters of L96 are a statistically complete sample
derived from the catalog of 732 nearby poor clusters of White et
al. (1999).  The clusters of the original sample were optically
selected by covering the entire sky north of $-3^{\circ}$ declination
and are identified as concentrations of 3 or more galaxies with
photographic magnitudes brighter than $15.7$ (from Zwicky et al.'s
1961-1968 catalogue), possessing a galaxy surface overdensity of
$21.5$.  The subsample of L96 is limited to the galactic latitude
range $|b|\ge 30^{\circ}$ and to the most dense and rich groups, i.e.
with 46.4 surface--density enhancement and with at least four Zwicky
galaxies.  L96 collected new redshifts and computed velocity
dispersions for several of these poor clusters.

The sample of ZM98 consists of 12 nearby optically--selected groups
from the literature (NED, NASA/IPAC Extragalactic Database) for which
there are existing, sometimes serendipitous, pointed PSPC observations
of the fields in which the groups lie. As pointed out by the authors,
this group sample is not representative of published group catalogs,
but is weighted towards X-ray groups.  By using multi-fiber
spectroscopy, ZM98 extend greatly the number of galaxies with
available redshift and present a sample of 1002 galaxy velocities.

Avoiding poor systems with $z\lesssim 0.01$, more strongly affected by
peculiar motions, and those which do not survive our procedure of
member selection (cf.~\S~3.1), we consider 36 poor clusters and 7 rich
groups for a total sample of 43 poor systems (hereafter PS) having
available magnitudes in COSMOS and/or APS (C-PS and A-PS samples,
respectively).  In particular, five poor systems have available
magnitudes in both photometric catalogs.

From ZM98 we take for each rich group: the group center, and galaxy
positions and redshifts, to apply our procedure of member selection
(cf. \S~3.1).  From L96 we take the mean velocity and the center for
each poor cluster: we use these data to collect galaxy positions and
redshifts within 1.5 \h from the cluster center by using NED, and then
apply our procedure of member selection.

\subsection{NOG Group Samples (HG and PG)}

We use the groups identified by Giuricin et al. (2000) in the NOG
sample. This is a complete, distance limited ($cz<6000$ \kss) and
magnitude limited ($B\leq 14$) sample of $\sim 7000$ optical galaxies,
which covers about $2/3$ of the sky $(|b| > 20^{\circ})$, and appears
to be quasi-complete in redshift ($97\%$). The authors identified the
groups by means of both the hierarchical and the percolation
``friend--of--friend'' methods: their final catalogs contain 475 and
513 groups, respectively (hereafter HG and PG).

From Giuricin et al. (2000) we take the data available for each
group's galaxy positions, redshifts, and corrected total blue
magnitudes.

\section{MASS AND LUMINOSITY ESTIMATES}

With the exception of the masses of CLs and luminosities of C-CLs, all
other mass and luminosity estimates are obtained in this study.  In
order to extend the work by G00 throughout this section, a great
effort is devoted to computing masses and luminosities in a consistent
way so as to obtain a large homogeneous sample of $M/L$ estimates.

In particular, G00 used mass and luminosity computed within the virial
radius $R_{vir}$, which defines, as usual in the context of cold dark
matter (CDM) cosmologies, the region where the matter overdensity is
$\sim 180$ for a $\Omega_m=1$ cosmology, or $\sim 350$ for a
$\Omega_m=0.3$ and $\Omega_{\Lambda}=0.7$ cosmology (which hereafter
we use as our reference model in this study).  In this region one can
assume a status of dynamical equilibrium and therefore reasonably
apply the virial theorem for the mass computation.

In our case, where we deal with systems spanning a large dynamical
range, to consider comparable physical regions is particularly useful
for mass and luminosity estimation.  The great advantage lies both in
considering a similar dynamical status (e.g., in the case of
variations among galaxy systems, cf. Zabludoff \& Mulchaey 1998b) and
a comparable galaxy population (in connection with color gradients,
e.g., Abraham et al. 1996). Therefore we compute both mass and
luminosity within $R_{vir}$.

\subsection{Mass Determination}

For poor systems, PCLs, we perform the same procedure already used by
G98, in particular with those recipes for poor data samples already
introduced by Girardi \& Mezzetti (2001).  Since the procedure was
already amply described in these works (cf. also Fadda et al. 1996),
here we only outline the main steps.

\subsubsection{Member Selection}

After having converted all galaxy velocities to galactocentric ones,
we perform the selection of member galaxies. First, we apply the
one-dimensional algorithm of the adaptive kernel technique by Pisani
(1993, see also Appendix A of Girardi et al. 1996) to find the
significant peaks in the velocity distribution.  The main cluster body
is generally identified as the highest significant peak. In some
particular cases, where the sampling is particularly poor, we had to
choose another peak in order to have a good coincidence (at least
within 1000 \kss) with the mean system velocity suggested by L96.  All
galaxies not belonging to the selected peak are rejected as non
cluster members.  F96 and G98 required that peaks must be significant
at the $99\%$ c.l., but in dealing with poor sampled systems we
follow the suggestion by Girardi \& Mezzetti (2001), considering
peaks having smaller significance $<99\%$ (but generally $>95\%$).  We
do not consider systems with multipeaked velocity distributions.

Afterwards, we use the combination of position and velocity
information to reveal the presence of surviving interlopers by
applying the ``shifting gapper'' (Fadda et al. 1996).  We reject
galaxies that are too far away in velocity (by $\ge1000$ \kss) from
the main body of galaxies at a given distance from the system center
(within a shifting annulus of 0.4 \h or large enough to include 15
galaxies). As for very poor samples with less than 15 members, we
reject galaxies that are too far away in velocity from the main body
of galaxies of the whole system.

At this point we recompute the system center for rich groups of ZM98
(by using the two--dimensional adaptive kernel method; cf. Pisani et
al. 1996; Girardi et al.  1996), while for the poor clusters of L96 we
retain the original centers, generally coming from a much larger
number of galaxies.

\subsubsection{Galaxy Velocity Dispersion}

We estimate the ``robust'' l.o.s. velocity dispersion, $\sigma_v$, by
using the biweight and the gapper estimators when the galaxy number is
larger or smaller than 15, respectively (cf. ROSTAT routines -- see
Beers Flynn, \& Gebhardt 1990), and applying the relativistic
correction and the usual correction for velocity errors (Danese, De
Zotti, \& di Tullio 1980).  For the poor clusters of L96, where
redshifts are taken from NED, we assume a typical velocity error of
$100$ \kss.

Following Fadda et al. (1996, cf. also Girardi et al. 1996) we analyze
the ``integral'' velocity dispersion profile (hereafter VDP), where
the dispersion at a given (projected) radius is evaluated by using all
the galaxies within that radius, i.e. $\sigma_v(<R)$. The VDPs make it
possible for us to check the robustness of the $\sigma_v$ estimate. In
fact, the presence of velocity anisotropy in galaxy orbits can
strongly influence the value of $\sigma_v$ computed for the central
cluster region, but does not affect the value of the $\sigma_v$
computed for the whole cluster (e.g., Merritt 1988). The VDPs of
nearby clusters show strongly increasing or decreasing behaviors in
the central cluster regions, but they are flattening out in the
external regions, suggesting that in such regions they are no longer
affected by velocity anisotropies.  Thus, while the $\sigma_v$-values
computed for the central cluster region could be a very poor estimate
of the depth of cluster potential wells, one can reasonably adopt the
$\sigma_v$ value computed by taking all the galaxies within the radius
at which the VDP becomes roughly constant.

When the data are good enough, also poor systems show an
(asymptotical) flatness in their VDPs (cf. Figure~1).  However, some
cases show a sharp increase towards the very external regions,
suggesting the presence of a neighboring system with a different mean
velocity (cf., e.g., A3391 and A3395 in Girardi et al. 1996). Since
the radius of our data samples is relatively large this situation is
not unexpected.  For these systems, we assume that the real system is
enclosed within the radius where the VDP sharply increases when there
are enough galaxies to detect a region of a flat profile (N79-298,
N79-283, N67-336, N45-363) or, otherwise, within the first galaxy
useful for computing $\sigma_v$ (S49-142, N67-317).

G98 computed the virial radius from: 
\begin{equation}
R_{vir}=[2\cdot\sigma_v/(1000~km~s^{-1})]~h^{-1}~Mpc,
\end{equation}
\noindent and we adopt the same definition.  Indeed, this is only a
first order approximation, but Girardi et al. (1998a) made a
recomputation for some choices of cosmological models: they found
that, for our reference model ($\Omega_m=0.3,~\Omega_{\Lambda}=0.7$),
the difference is $\sim 10\%$ for $R_{vir}$ and $\sim 5\%$ for the
mass within $R_{vir}$. Since we are interested in fixing a radius for
dealing with consistent physical regions in different systems, rather
than with precise cosmological computations, this choice is well
suited to our aims.

\clearpage

\end{multicols}
%\begin{figure}
\includegraphics{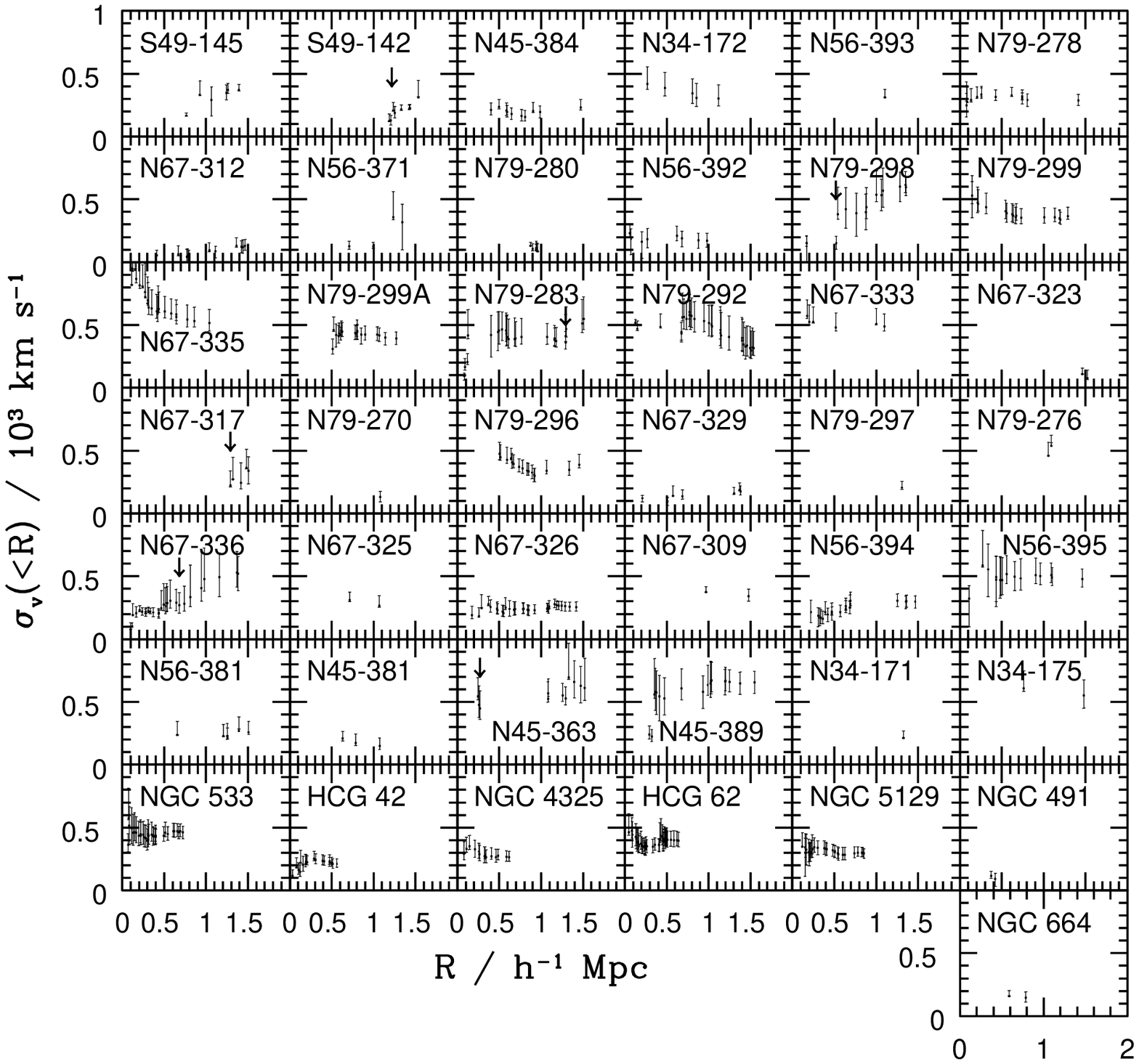}
%\special{psfile=f2.eps angle=0 voffset=-100 hoffset=-0 vscale=60 hscale=60}
$\ \ \ \ \ \ $\\
\vspace{15truecm}
$\ \ \ $\\
%\vspace{1cm}
{\small\parindent=3.5mm {Fig.}~1.---
Integrated line--of--sight velocity dispersion
profiles $\sigma_v(<R)$, where the dispersion at a given (projected)
radius from the system center is estimated by considering all
galaxies within that radius. The bootstrap error bands at the $68\%$
c.l. are shown. The arrows indicate where six clusters 
are truncated according to the analysis of velocity
dispersion profiles (see text).}
\vspace{5mm}
\begin{multicols}{2}
%%%%%%%%%%%%%%

\subsubsection{Galaxy Spatial Distribution}
  
In order to estimate the virial mass, one must compute the radius
appearing in the virial theorem (Limber \& Mathiews 1960), which is,
indeed, larger than the harmonic radius by about a factor of two.  In
particular, since we want to compute the mass within $R_{vir}$, only
the $N$ galaxies within $R_{vir}$ are considered.  Here we use the
luminosity-unweighted, projected version of this radius, $R_{PV}$
(cf. Giuricin, Mardirossian, \& Mezzetti 1982; please note that G98
referred to $R_{PV}$ as ``virial radius''):
\begin{equation}
R_{PV}=N(N-1)/(\Sigma_{i> j} R_{ij}^{-1}),
\end{equation}
\noindent where $R_{ij}$ are the projected mutual galaxy
distances. 

Unfortunately, for several poor systems the number of galaxies within
$R_{vir}$ is very small ($N\le6$) or the sampled region is smaller
than $R_{vir}$ (26 systems). In these cases the estimate of $R_{PV}$
can be recovered in an alternative way from the knowledge of the
galaxy surface density profile $\Sigma(R)$: Girardi et al. (1995)
presented this method for the King--like distribution
$\Sigma (R)={\Sigma (0)}/{[1+(R/R_c)^2]^{\alpha}}$, where $R_c$ is the
core radius and $\alpha$ is the parameter which describes the galaxy
distribution in external regions (cf. the $\beta$-profile used in
X-ray surface brightness analyses).

The alternative procedure described by Girardi et al. (1995) allows
one to compute $R_{PV}$ at each radius and, in particular, we compute
$R_{PV}$ at $R_{vir}$. Here we use the same parameters as the
King--like distribution already used by G98: $\alpha=0.7$ and
$R_c/R_{vir}=0.05$.  In fact, by using the 17 systems sampled with
$N\leq 10$ galaxies within $R_{vir}$ and the same Maximum Likelihood
approach, we fit consistent values. We find:
$\alpha=0.64_{-0.02}^{+0.07}$ (median value and $90\%$ errors), which
corresponds to a galaxy volume--density of $\rho \propto r^{-2.4}$ for
$R>>R_C$; and $R_c/R_{vir}=0.03_{-0.01}^{+0.03}$.

\subsubsection{Virial Mass}

Assuming that clusters are spherical, non--rotating systems, and that
the internal mass distribution follows galaxy distribution, cluster  
masses can be computed with the virial theorem (e.g., Limber \&
Mathiews 1960; The \& White 1986) as:
\begin{equation}
M=M_V-C=\frac{3\pi}{2} \cdot \frac{\sigma_v^2 R_{PV}}{G}-C,
\end{equation}
\noindent where the $C$ correction takes into account
that the system is not fully enclosed within 
the boundary radius, $b$,
here $R_{vir}$. The correction can be written as:
\begin{equation}
C=M_V \cdot 4 \pi b^3 \frac{\rho(b)}{\int^{b}_0 4\pi r^2\rho dr}\left[\sigma_r(b)/\sigma(<b)\right]^2,
\end{equation}
\noindent and requires knowledge of the velocity anisotropy of
galaxy orbits.  In fact, $\sigma_r(b)$ is the radial component of the
velocity dispersion $\sigma(b)$, while $\sigma(<b)$ refers to the
integrated velocity dispersion within $b$; here $b=R_{vir}$.  

In order to give the $C$--correction for each individual cluster, G98
used a profile indicator, $I_p$, which is the ratio between
$\sigma_v(<0.2\times R_{vir})$, the l.o.s. velocity dispersion
computed by considering the galaxies within the central cluster region
of radius $R=0.2\times R_{vir}$, and the global $\sigma_v$. According
to the values of this parameter, they divided clusters into three
classes: ``A'' clusters with a decreasing profile ($I_p>1.16$), ``C''
clusters with an increasing profile ($I_p<0.97$), and an intermediate
class ``B'' of clusters with very flat profiles ($0.97<I_p<1.16$).
Each of the three types of profiles can be explained by models with a
different kind of velocity anisotropy: fully isotropical (B), with a
radial component only in external regions (A), or with a circular
component in central regions (C); cf. Figure~3 of G98 and relative
comments.

In the same way, we can define 6, 4, and 5 systems belonging to class
A, B, and C, respectively, and for each class we use the respective
value $[\sigma_r(R_{vir})/\sigma(<R_{vir})]^2$ and typical galaxy
distribution $\rho(r)$ to determine the $C$--corrections.  For systems
where we cannot define the type of profile we assume isotropical
velocities, i.e. the type ``B'', already shown to be the most adequate
for describing clusters and acceptable also for poor clusters (at the
$39\%~\chi^2$ probability). Figure~2 compares the observational
velocity dispersion profile $\sigma_v(R)$, as computed by combining
together the galaxies of all 17 ``well--sampled'' systems, to the
three models with different velocity anisotropy recovered by using the
Jeans equation: both the model with isotropical velocity and that with
circular velocity anisotropy are acceptable.

The median value of the correction of the sample is $20\%$, similar to
that of G98, and also to that of CNOC clusters (Calberg, Yee, \&
Ellingson 1997).
 
Table~2 lists the results of the dynamical analysis: the system name,
Col.~(1); $N_f$, the number of galaxies with measured redshift in each
cluster field [Col.~(2)]; $N_m$, the number of member galaxies after
checking with VDP [Col.~(3)] and used to compute $V$, the mean
velocity, [Col.~(4)] and $\sigma_v$, the global l.o.s. velocity
dispersion, with the respective bootstrap errors [Col.~(5)]: only
systems with $N_m\ge 5$ are retained in the sample; $R_{vir}$, the
virial radius which defines the region of dynamical equilibrium
[Col.~(6)]; $N$, the number of member galaxies within $R_{vir}$
[Col.~(7)]; $R_{PV}$, the projected radius used in the virial theorem
and here computed within $R_{vir}$, with the respective jacknife error
(a $25\%$ error is assumed for radii computed from the alternative
theoretical formula; cf. G98) [Col.~(8)]; $T$, the type of velocity
dispersion profile [Col.~(9)]; $M$, the virial mass contained within
$R_{vir}$ after pressure surface term correction with the
corresponding errors [Col.~(10)].  The percent errors on $M$ are the
same as for $M_V$, i.e. we take into account the errors on $\sigma_v$
and $R_{PV}$ and neglect the uncertainties on $C$--correction.

The median percent error on mass is $\sim 40\%$, but varies
with the mass (see also Girardi et al. 1998a), ranging  from $\sim 30\%$
for massive systems, $M>5\times 10^{14}~M_{\odot}$, to $\sim 75\%$
for less massive systems, $M< 5 \times 10^{13}$ $M_{\odot}$.

%\end{multicols}
%\begin{figure}
\includegraphics{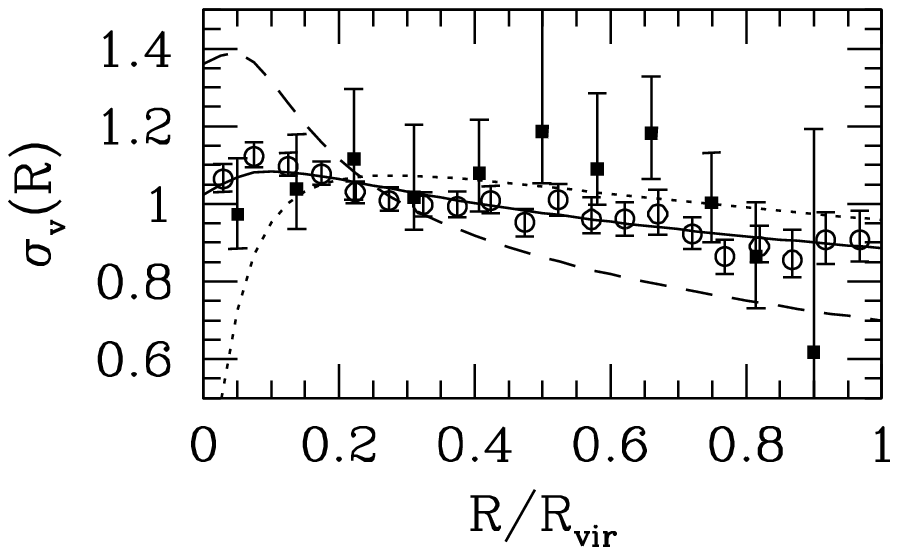}
%\special{psfile=f2.eps angle=0 voffset=-100 hoffset=-0 vscale=60 hscale=60}
$\ \ \ \ \ \ $\\
\vspace{5truecm}
$\ \ \ $\\
%\vspace{1cm}
{\small\parindent=3.5mm {Fig.}~2.---
The (normalized) line--of--sight velocity
dispersion, $\sigma_v(R)$, as a function of the (normalized) projected
distance from the system center.  The points represent data combined
from all systems and binned in equispatial intervals. We give the
robust estimates of velocity dispersion and the respective bootstrap
errors.  We give the results for poor systems (open circles) and for
nearby clusters taken from Girardi et al. (1998b, filled circles).
The solid, dotted, and dashed lines represent models with different
kinds of velocity anisotropy: isotropic, circular in central regions,
and radial in external regions, respectively (see text).}
\vspace{5mm}
%\begin{multicols}{2}
%%%%%%%%%%%%%%

Finally, we note that four poor clusters here analyzed were already
studied by G98: N34-172/MKW1, N67-312/MKW10, N67-335/MKW4,
N67-336/MKW12. The two mass estimates, based on partially different
data samples (G98 considered only homogeneous data from specific cluster
studies), show fair agreement.

\end{multicols}
%%TAB2%\vspace{-6mm}
\hspace{-13mm}
\begin{minipage}{20cm}
\renewcommand{\arraystretch}{1.2}
\renewcommand{\tabcolsep}{1.2mm}
\begin{center}%\vspace{-3mm}
TABLE 2\\%\vspace{2mm}
{\sc Results of Dynamical Analysis \\}
\footnotesize 
%%%%%%%%%%%%%%%%%%%%%%%%%%%%%%%%%%%%%%%%%%%%%%%%%%%%%%%%%%%%%%%%%%%%%%%
%
%                        TABLE 2
%
%%%%%%%%%%%%%%%%%%%%%%%%%%%%%%%%%%%%%%%%%%%%%%%%%%%%%%%%%%%%%%%%%%%%%%%
\begin{tabular}{lrrrrrrrcr}
\hline \hline
\multicolumn{1}{c}{Name}
&\multicolumn{1}{c}{$N_f$}
&\multicolumn{1}{c}{$N_m$}
&\multicolumn{1}{c}{$V$}
&\multicolumn{1}{c}{$\sigma_v$}
&\multicolumn{1}{c}{$R_{vir}$}
&\multicolumn{1}{c}{$N$}
&\multicolumn{1}{c}{$R_{PV}$\tablenotemark{a}}
&\multicolumn{1}{c}{T}
&\multicolumn{1}{c}{$M$}
\\
\multicolumn{1}{c}{}
&\multicolumn{1}{c}{}
&\multicolumn{1}{c}{}
&\multicolumn{1}{c}{$km\,s^{-1}$}
&\multicolumn{1}{c}{$km\,s^{-1}$}
&\multicolumn{1}{c}{$h^{-1}\,Mpc$}
&\multicolumn{1}{c}{}
&\multicolumn{1}{c}{$h^{-1}\,Mpc$}
&\multicolumn{1}{c}{}
&\multicolumn{1}{c}{$h^{-1}\,10^{14}M_{\odot}$}
\\
\multicolumn{1}{c}{(1)} 
&\multicolumn{1}{c}{(2)}
&\multicolumn{1}{c}{(3)}
&\multicolumn{1}{c}{(4)}
&\multicolumn{1}{c}{(5)}
&\multicolumn{1}{c}{(6)}
&\multicolumn{1}{c}{(7)}
&\multicolumn{1}{c}{(8)}
&\multicolumn{1}{c}{(9)}
&\multicolumn{1}{c}{(10)}
\\ 
\hline
S49-145        & 29& 10&  6812&  371$^{+   57}_{-   72}$& .74&  4& (.55$\pm$ .14)& - &  .66$^{+  .26}_{-  .31}$\\
S49-142        & 20&  5&  6299&  122$^{+  140}_{-  140}$& .24&  3& (.18$\pm$ .05)& - &  .02$^{+  .05}_{-  .05}$\\
N45-384        & 26& 14&  7833&  232$^{+   50}_{-   82}$& .46&  5& (.35$\pm$ .09)& - &  .16$^{+  .08}_{-  .12}$\\
N34-172        & 17&  9&  6007&  302$^{+   89}_{-   82}$& .60&  6& (.45$\pm$ .11)& - &  .36$^{+  .23}_{-  .21}$\\
N56-393        & 24&  5&  6663&  311$^{+  174}_{-   27}$& .62&  4& (.46$\pm$ .12)& - &  .39$^{+  .45}_{-  .12}$\\
N79-278        & 40& 17&  8983&  287$^{+   37}_{-   59}$& .57& 12& .25$\pm$ .06& C &  .19$^{+  .07}_{-  .09}$\\
N67-312        & 33& 16&  6036&  131$^{+    9}_{-   67}$& .26&  3& (.19$\pm$ .05)& - &  .03$^{+  .01}_{-  .03}$\\
N56-371        & 53&  8&  8130&  319$^{+   20}_{-  101}$& .64&  4& (.47$\pm$ .12)& - &  .42$^{+  .12}_{-  .29}$\\
N79-280        & 17& 10&  9390&   87$^{+   10}_{-   80}$& .17&  1& (.13$\pm$ .03)& - &  .01$^{+  .00}_{-  .02}$\\
N56-392        &174& 12&  7995&  172$^{+   15}_{-   95}$& .34&  8& .21$\pm$ .07& A &  .04$^{+  .01}_{-  .04}$\\
N79-298        & 46&  7&  4478&  152$^{+   47}_{-  141}$& .30&  6& (.23$\pm$ .06)& - &  .05$^{+  .03}_{-  .09}$\\
N79-299B       & 26& 22&  6920&  364$^{+   46}_{-   54}$& .73& 15& .58$\pm$ .14& A &  .47$^{+  .16}_{-  .18}$\\
N67-335        & 40& 27&  5922&  515$^{+  104}_{-   87}$&1.03& 26& .71$\pm$ .11& A & 1.14$^{+  .49}_{-  .42}$\\
N79-299A       & 26& 22&  6910&  391$^{+   40}_{-   60}$& .78& 12& .47$\pm$ .14& - &  .63$^{+  .22}_{-  .27}$\\
N79-283        & 52& 24&  7970&  360$^{+   84}_{-   48}$& .72& 18& .55$\pm$ .13& B &  .62$^{+  .32}_{-  .22}$\\
N79-292        &116& 34&  7290&  318$^{+   88}_{-   76}$& .64&  7& .48$\pm$ .09& A &  .29$^{+  .17}_{-  .15}$\\
N67-333        & 20& 10& 14176&  489$^{+  102}_{-   47}$& .98&  8& .37$\pm$ .11& A &  .53$^{+  .28}_{-  .19}$\\
N67-323        & 15&  6&  9212&   79$^{+   43}_{-   51}$& .16&  2& (.12$\pm$ .03)& - &  .01$^{+  .01}_{-  .01}$\\
N67-317        & 35&  5&  7021&  223$^{+  253}_{-  253}$& .45&  1& (.33$\pm$ .08)& - &  .14$^{+  .33}_{-  .33}$\\
N79-270        &  9&  5&  6746&  130$^{+   73}_{-   70}$& .26&  3& (.19$\pm$ .05)& - &  .03$^{+  .03}_{-  .03}$\\
N79-296        & 33& 20&  6779&  388$^{+   69}_{-   62}$& .78& 11& .78$\pm$ .18& - & 1.03$^{+  .43}_{-  .40}$\\
N67-329        & 13& 11&  6842&  176$^{+   24}_{-   94}$& .35&  5& (.26$\pm$ .07)& - &  .07$^{+  .03}_{-  .08}$\\
N79-297        & 11&  5&  8730&  206$^{+   76}_{-  136}$& .41&  2& (.31$\pm$ .08)& - &  .11$^{+  .09}_{-  .15}$\\
N79-276        & 36&  6& 11059&  540$^{+  148}_{-   70}$&1.08&  5& (.80$\pm$ .20)& - & 2.04$^{+ 1.23}_{-  .74}$\\
N67-336        & 71& 25&  5843&  270$^{+   65}_{-   57}$& .54& 20& .45$\pm$ .06& C &  .31$^{+  .15}_{-  .14}$\\
N67-325        &  8&  6&  5184&  265$^{+   56}_{-  226}$& .53&  4& (.39$\pm$ .10)& - &  .24$^{+  .12}_{-  .42}$\\
N67-326        &146& 33&  4532&  256$^{+   26}_{-   50}$& .51& 11& .48$\pm$ .13& - &  .27$^{+  .09}_{-  .13}$\\
N67-309        & 16&  6&  8057&  340$^{+  108}_{-   33}$& .68&  4& (.51$\pm$ .13)& - &  .51$^{+  .35}_{-  .16}$\\
N56-394        & 49& 23&  8586&  296$^{+   35}_{-   59}$& .59& 14& .72$\pm$ .10& - &  .55$^{+  .15}_{-  .23}$\\
N56-395        & 27& 20&  8137&  477$^{+   82}_{-   61}$& .95& 16& .67$\pm$ .18& C & 1.44$^{+  .62}_{-  .53}$\\
N56-381        & 16& 10&  8950&  259$^{+   53}_{-   79}$& .52&  4& (.39$\pm$ .10)& - &  .23$^{+  .11}_{-  .15}$\\
N45-381        & 50&  7& 11198&  150$^{+   91}_{-   91}$& .30&  1& (.22$\pm$ .06)& - &  .04$^{+  .05}_{-  .05}$\\
N45-363        & 26&  7& 10995&  449$^{+  146}_{-   81}$& .90&  7& (.67$\pm$ .17)& - & 1.17$^{+  .82}_{-  .52}$\\
N45-389        & 24& 22&  9522&  656$^{+   75}_{-   84}$&1.31& 20&1.18$\pm$ .19& B & 4.43$^{+ 1.25}_{- 1.35}$\\
N34-171        & 13&  5&  5484&  215$^{+   79}_{-   75}$& .43&  2& (.32$\pm$ .08)& - &  .13$^{+  .10}_{-  .10}$\\
N34-175        & 12&  6&  8894&  550$^{+  212}_{-   76}$&1.10&  5& (.82$\pm$ .20)& - & 1.50$^{+ 1.22}_{-  .56}$\\
NGC 533        & 99& 36&  5518&  464$^{+   54}_{-   47}$& .93& 36& (.69$\pm$ .17)& A & 1.30$^{+  .44}_{-  .42}$\\
HCG 42         &106& 22&  3719&  211$^{+   22}_{-   43}$& .42& 17& .32$\pm$ .06& B &  .13$^{+  .04}_{-  .06}$\\
NGC 4325       & 68& 18&  7468&  265$^{+   36}_{-   40}$& .53& 16& .50$\pm$ .10& C &  .31$^{+  .10}_{-  .11}$\\
HCG 62         &106& 46&  4307&  396$^{+   57}_{-   55}$& .79& 46& (.59$\pm$ .15)& B &  .87$^{+  .33}_{-  .33}$\\
NGC 5129       & 85& 33&  6938&  294$^{+   33}_{-   38}$& .59& 26& .56$\pm$ .06& C &  .42$^{+  .11}_{-  .12}$\\
NGC 491        &104&  6&  3987&   92$^{+   70}_{-   38}$& .18&  4& (.14$\pm$ .03)& - &  .01$^{+  .02}_{-  .01}$\\
NGC 664        & 67&  6&  5489&  148$^{+   90}_{-   79}$& .30&  4& (.22$\pm$ .06)& - &  .04$^{+  .05}_{-  .05}$\\
\hline
\end{tabular}

\end{center}
\vspace{-2mm}
{\footnotesize\parindent=3mm 
$^a$~Values in brackets are those computed 
through the alternative estimate by using the typical galaxy
distribution (see text).}

\end{minipage}
\begin{multicols}{2}

\subsection{Luminosity Determination}

\subsubsection{COSMOS and APS Catalogs} 

Following G00, we derive from the COSMOS catalogue the magnitudes for
8 poor systems of the PS sample.

The COSMOS catalogue is described by Yentis et al. (1992), and a part,
the Edinburgh--Durham Southern Galaxy Catalogue (EDSG) is well
analyzed by Heydon-Dumbleton, Collins, \& MacGillivray (1989).  The
EDSGC is nominally quasi--complete to $B_j=20$. A more conservative
limiting apparent magnitude was suggested by Valotto et al. (1997),
who found that galaxy counts follow a uniform law for
$B_j<19.4$. Accordingly, and following G00, we decide to adopt a
limiting magnitude of $B_j=19.4$ for the COSMOS catalog.  Within this
limit, we adopt a completeness value of $91\%$ for areas containing
galaxy systems, suggested as being more appropriate for areas of high
surface density (Katgert et al. 1998) and the nominal $95\%$
completeness value for the rest of the catalog (Heydon-Dumbleton et
al. 1989). These levels of incompleteness are assumed for all
magnitudes down to $B_j=19.4$, since Katgert et al. (1998) found that
the magnitude distribution of missing galaxies is essentially the same
as for sampled galaxies.  Moreover, we note that, although COSMOS
magnitudes are isophotal magnitudes, the threshold on average being
only $8\%$ above the sky (Heydon-Dumbleton, Collins, \& MacGillivray
1988) the difference between COSMOS magnitudes and ``total''
magnitudes becomes significant only at the faint limit (well below our
limiting magnitude, cf. Shanks, Stevenson, \& Fong 1984).

For each system of C-PS we select galaxies with magnitudes $B_j<19.4$
within circular regions with a radius equal to $R_{vir}$ and a center
as chosen in \S~3.1.1.

Moreover, we take magnitudes for 52 CLs and 40 PSs
from the fields which are currently on--line from the APS
Revised Catalog of POSS I.

The APS Catalog is the result of scans of glass duplicates of the blue
(O) and red (E) plates of the original Palomar Observatory Sky Survey
(POSS I) for all 664 fields with $|b| > 20^{\circ}$.  The operation of
the Automated Plate Scanner (APS) and the scanning procedures and
parameters are described in detail in Pennington et al. (1993).  Here
we consider $B_{APS}$ band--magnitude, corresponding to the (O) Sky
Survey plates, which is an isophotal magnitude within the level of
surface brightness of $\mu \sim 26.5$ mag per square arcsec.
Comparisons with available photometry in the RC3 and for fainter
galaxies at the North Galactic Pole find that APS--derived integrated
galaxy magnitudes show no systematic photometric errors and a typical
rms scatter of 0.2 to 0.3 magnitudes (Odewahn \& Aldering 1995).
Several checks suggest that the catalog is quasi--complete for
$B_{APS}<19$--$20$ mag (Odewahn et al. 1993; Odewahn \& Aldering
1995). We assume a completeness of $85\%$ down to 19.4 $B_{APS}$ as
recovered by Odewahn \& Aldering (1995) from a comparison with a
previous photometric catalog of Coma. In particular, we assume the
same level of incompleteness for all magnitudes down to $B_j=19.4$;
this seems true at least for $B_{APS}>15$ (cf. Figure~11 by Odewahn et
al. 1993).

For each system of A-CL and A-PS we select galaxies with magnitudes
$B_{APS}<19.4$ within circular regions with a radius equal to
$R_{vir}$ and a center as chosen in \S~3.1.1 for poor systems, or
computed by G00 for clusters.  Since the APS catalog is still in
progress, before including a system in our study we have checked by
visual inspection that the photometric data fully cover the selected
system region.

\end{multicols}
%\begin{figure}
\includegraphics{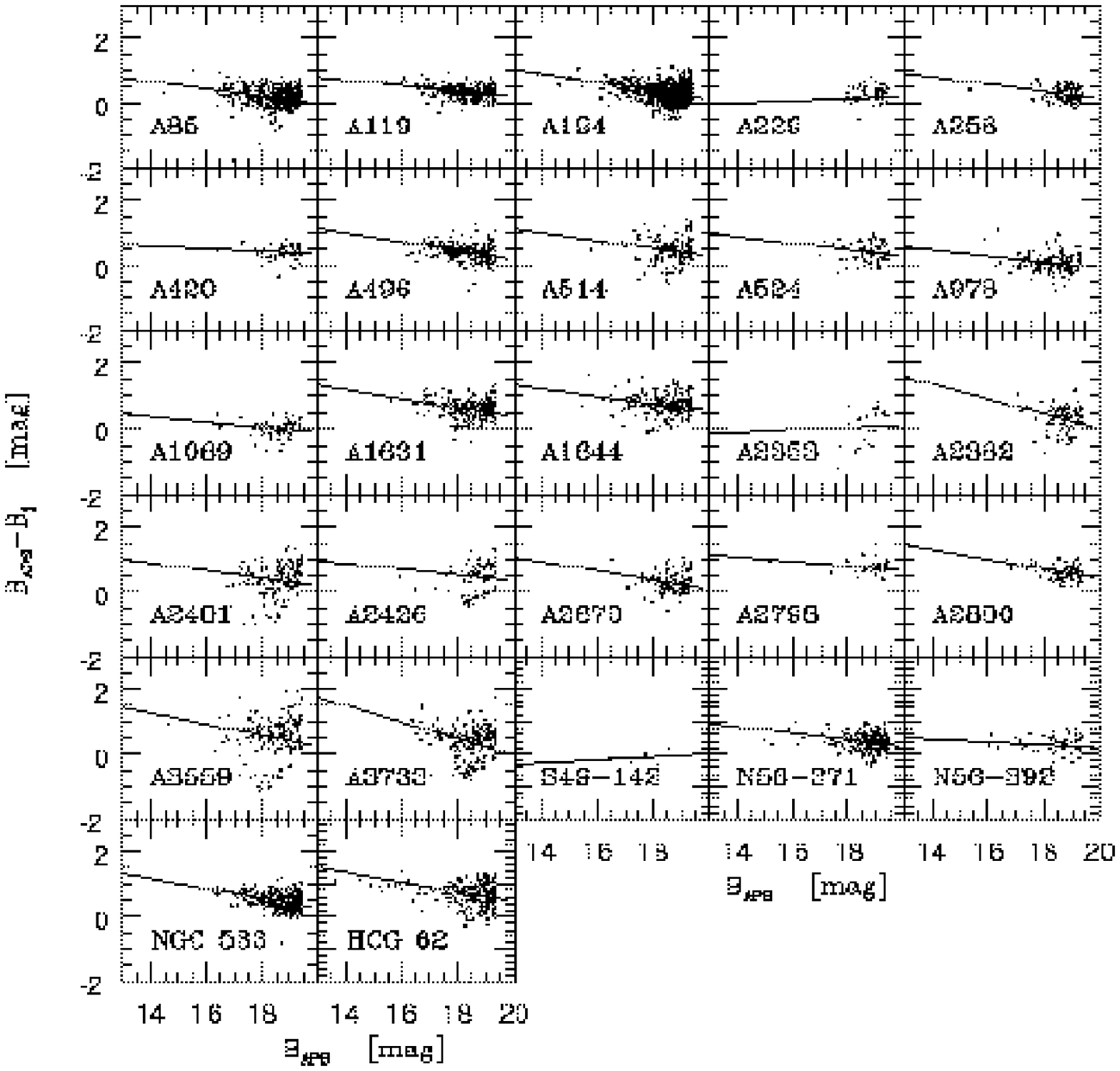}
%\special{psfile=f2.eps angle=0 voffset=-100 hoffset=-0 vscale=60 hscale=60}
$\ \ \ \ \ \ $\\
\vspace{12truecm}
$\ \ \ $\\
%\vspace{1cm}
{\small\parindent=3.5mm {Fig.}~3.---
Comparison
of magnitudes from COSMOS and APS catalogs
for the galaxies of the 27 systems having both photometries
available. The faint lines represent the result of the
(bisecting) fitting of $B_j$ vs. $B_{APS}$.
}
\vspace{5mm}
\begin{multicols}{2}
%%%%%%%%%%%%%%

\subsubsection{Magnitude Conversion and Correction} 

In order to homogenize the photometry coming from the two catalogs, as
well as for comparison with results by G00, we convert $B_{APS}$
magnitudes into the $B_j$ band.  We consider the 27 systems for which
both photometries are available: 22 A-CLs in common with C-CLs of G00,
and five PSs.  On the base of similar positions, with a maximum
distance of $0.1'$, we select 4845 system galaxies having magnitudes
in both catalogs (note that for this analysis we consider somewhat
larger areas).  We fit $B_j$ vs. $B_{APS}$ magnitudes by using the
unweighted bisecting procedure (Isobe et al. 1990) for each of the 27
systems; cf.  Figure~3 where we show $B_{APS}-B_j$ vs. $B_{APS}$
magnitudes for the sake of clarity.

We verify that slopes (and intercepts) of all straight lines can be
derived from only one parent value, according to the Homogeneity test
(or Variance--ratio test, cf. e.g., Guest 1961), and combine together
data of all 27 systems. Figure~4 shows the cumulative relation between
the two magnitude bands and the corresponding quadratic fit (via the
MINUIT subroutine of CERN Libraries):
\begin{equation}
B_{APS}-B_j=0.23+0.19\cdot B_{APS}-1.02\cdot 10^2\cdot B_{APS}^2,
\end{equation}
\noindent 
performed by using the 1021 galaxies with $B_{APS}<18~mag$ in order
to avoid the bias on the difference $(B_{APS}-B_j)$ caused by the limit
in $B_j$.

%\end{multicols}
%\begin{figure}
%\special{psfile=f4.eps angle=0 voffset=-570 hoffset=5 vscale=60 hscale=60}
\includegraphics{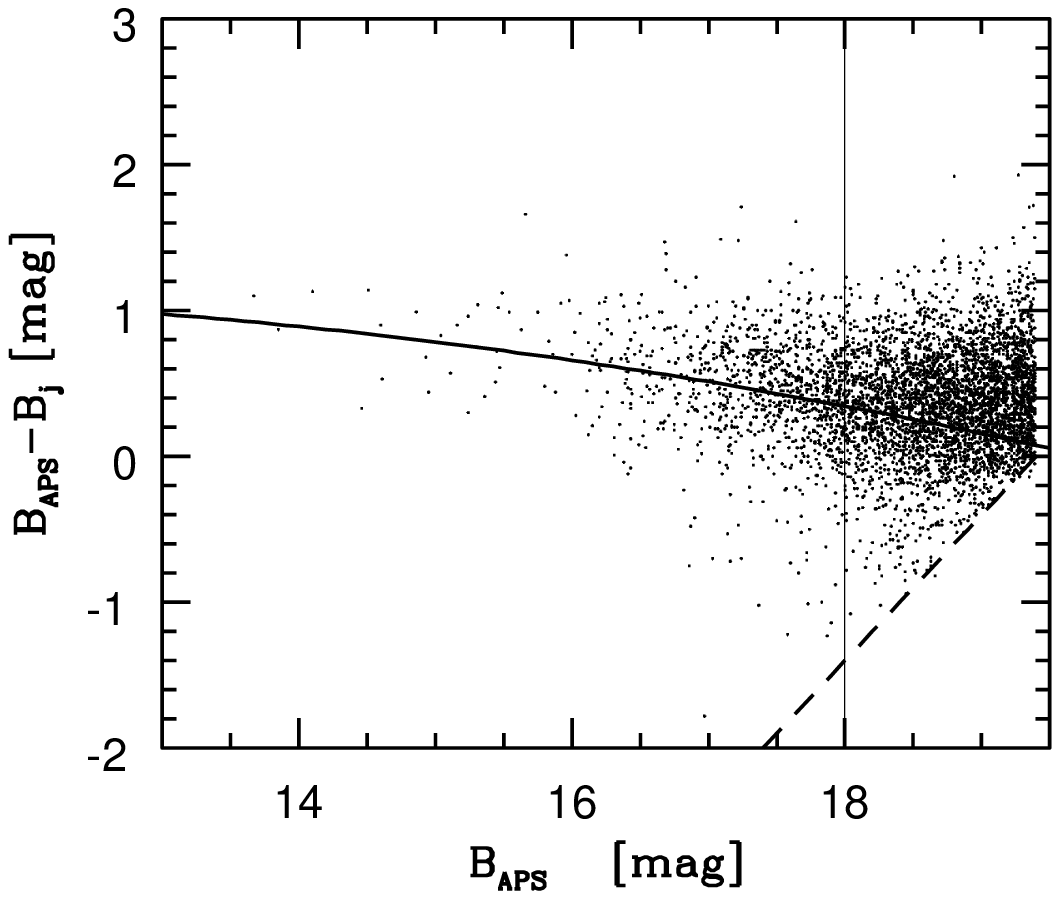}
$\ \ \ \ \ \ $\\
\vspace{6truecm}
$\ \ \ $\\
%\vspace{1cm}
{\small\parindent=3.5mm {Fig.}~4.---
The relationship between COSMOS and APS
magnitudes obtained by combining data for all 27 systems shown in
Figure~3.  The solid line represents the best quadratic fit, which is
performed on the data with $B_{APS}<18$ mag to avoid the bias on the
difference ($B_{APS}-B_j$) caused by the limit in $B_j$, as shown by
the dashed line.
}
\vspace{5mm}
%\begin{multicols}{2}

After having homogeneized all magnitudes to COSMOS $B_j$, following
G00 we apply the correction of Lumsden et al. (1997) to convert COSMOS
$B_j$ magnitudes to the CCD--based magnitude scale. This correction is
consistent with our choice to use the luminosity function and counts
by Lumsden et al. (1997).  Finally, we correct each galaxy magnitude
for (1)~Galactic absorption by assuming the absorption in the blue
band as given by de Vaucouleurs et al. (1991) in the region of each
system, and (2) $K$--dimming by assuming that all galaxies lie at the
average system redshift (Colless 1989).  The correction for internal
galactic absorption will be taken into account at the end of the
procedure to estimate luminosities.

\subsubsection{Luminosities from COSMOS and APS}

For all C-PSs, A-PSs and A-CLs we compute total luminosities within
$R_{vir}$ following the procedure already outlined for C-CLs by G00
who used COSMOS.  Having already homogeneized the original APS magnitudes
to the $B_j$ band, the whole procedure is the same except for the value of
the magnitude incompleteness in the catalogs (cf. point 2 below). Here
we give a short summary of the procedure.

1)~We compute observed system luminosities, $L_{B_j,obs}$, obtained
by summing the individual absolute luminosities of all galaxies and
assuming $B_{j,\odot}=5.33$ (i.e., $B_{\odot}=5.48$ and using the
conversion to $B_j$ by Kron 1978).

2)~For samples from the COSMOS (or APS) catalog we correct for $91\%$
(or $85\%$) incompleteness, obtaining
$L_{B_j,compl}=L_{B_j,obs}/0.91$.  (or $L_{B_j,compl}=L_{B_j,obs}/0.85$),
and the same correction is also applied to number counts.

3)~We subtract the average fore/background luminosity obtained from
the mean field $B_j$ counts by Lumsden et al. (1997) using the EDSGC
(cf. their Figure~2, corrected for  $95\%$ completeness):
$L_{B_j,corr}=L_{B_j,compl}-L_{B_j,meanfield}$. We correct counts in a
similar way.  The median correction is $29\%$.

4)We include the luminosity of faint galaxies below the magnitude
completeness limit $L_{B_j,extr}=L_{B_j,corr}+L_{B_j,faint}$, where
$L_{B_j,faint}$ is obtained by extrapolating the usual Schechter
(1976) form for the cluster luminosity function (with
$M_{Bj}^*=-20.16$ for the characteristic magnitude and $\alpha=-1.22$
for the slope as determined by Lumsden et al. 1997), normalized by
using the observed (corrected) galaxy number counts for $-21 \le
M_{B_j} \le -18$ (cf. eqs.~1 and 2 of G00).  The median correction is
$5\%$.

5)~The final luminosity estimate of galaxy systems, $L_{B_j}$, takes
into account the internal galactic absorption by adopting a correction
of $\Delta B_j=0.1$ mag.  On account of to this correction, the
luminosity increases by about $10\%$.

$L_{B_j,c,COSMOS}$ and $L_{B_j,c,APS}$ denote luminosities
computed for the two catalogs.

The above method for the fore/background correction does not take into
account the local field--to--field count variations, which lead to
random errors.  Since this correction is the largest one in our
procedure, it is worthwhile considering an alternative procedure, too.
Following G00, we also consider the method recently used by Rauzy,
Adami, \& Mazure (1998) to take into account the very local field,
i.e.  the presence of a nearby group along the system l.o.s.. This
method is based on the idea that having redshifts for all galaxies
would allow an unambiguous determination of the membership and thus
the solution of the fore/background correction.  Rauzy et al.
suggested correcting cluster luminosity (and counts) by assuming that
the fraction of members of the examined photometric sample
corresponds, for luminosity and number, to that computed on a
corresponding sample of galaxies all having redshift
($L_{B_j,corr}=L_{B_j,obs}*f_L$ and $N_{corr}=N_{obs}*f_N$).  The
drawback of this method is obviously that the estimated member
fraction computed in the redshift sample depends on the magnitude
limit and the extension of the sampled region.  Here we use the
redshift samples analyzed in \S~3.1 for poor systems and those
analyzed by G98 for clusters in order to compute $f_N$ and $f_L$
within $R_{vir}$.  The fraction $f_N$ is directly recovered by
comparing the number of member galaxies with those in the field.  As
for clusters, following G00, we compute $f_L$ when original redshift
samples have available magnitudes, or assume that $f_L=f_N$ when
luminosities are not directly available (since $f_L\sim f_N$).  As for
poor systems, we compute $f_L$ after assigning to galaxies of redshift
samples the corresponding magnitudes taken from the photometric
samples (here $f_L>f_N$ for a typical factor of $30\%$); when less
than five galaxies are available we use the median values $f_L=0.8$
and $f_N=0.6$ as computed on all better sampled poor
systems. $L_{B_j,f,COSMOS}$ and $L_{B_j,f,APS}$ denote our alternative
luminosity estimates.

Table~3 gives the results of luminosity computation, reporting the
results of G00 too, for both clusters and poor systems (CL+PS sample,
162 systems), which make up a very homogeneous sample for both mass and
luminosity estimates. We list: the system name [Col.~(1)]; the adopted
system center [Col.~(2)]; $R_{vir}$, the virial radius [Col.~(3)];
$N_{APS}$, the number of galaxies within $R_{vir}$, used to compute
$L_{Bj,c,APS}$ and $L_{Bj,f,APS}$, the two alternative APS
luminosities based on two procedures for the fore/background
correction [Cols.~(4), and (5), respectively]; $N_{COSMOS}$, the
number of galaxies within $R_{vir}$, used to compute $L_{Bj,c,COSMOS}$
and $L_{Bj,f,COSMOS}$, the two alternative COSMOS luminosities
[Cols.~(6), and (7), respectively].

\end{multicols}
%%TAB3%\vspace{-6mm}
\hspace{-13mm}
\begin{minipage}{20cm}
\renewcommand{\arraystretch}{1.2}
\renewcommand{\tabcolsep}{1.2mm}
\begin{center}%\vspace{-3mm}
TABLE 3\\%\vspace{2mm}
{\sc Luminosity Estimates \\}
\footnotesize 
%%%%%%%%%%%%%%%%%%%%%%%%%%%%%%%%%%%%%%%%%%%%%%%%%%%%%%%%%%%%%%%%%%%%%%%
%
%                        TABLE 2
%
%%%%%%%%%%%%%%%%%%%%%%%%%%%%%%%%%%%%%%%%%%%%%%%%%%%%%%%%%%%%%%%%%%%%%%%
%
%                        TABLE 3
%
%%%%%%%%%%%%%%%%%%%%%%%%%%%%%%%%%%%%%%%%%%%%%%%%%%%%%%%%%%%%%%%%%%%%%%%
\begin{tabular}{lcrrcrc}
\hline \hline
\multicolumn{1}{c}{Name}
&\multicolumn{1}{c}{Center}
&\multicolumn{1}{c}{$R_{vir}$}
&\multicolumn{1}{c}{$N_{APS}$}
&\multicolumn{1}{c}{$L_{B_j,APS}$\tablenotemark{a}}
&\multicolumn{1}{c}{$N_{COSMOS}$}
&\multicolumn{1}{c}{$L_{B_j,COSMOS}$\tablenotemark{a,b}}
\\
\multicolumn{1}{c}{}
&\multicolumn{1}{c}{$\alpha(2000)$-$\delta(2000)$}
&\multicolumn{1}{c}{$h^{-1}\,Mpc$}
&\multicolumn{1}{c}{}
&\multicolumn{1}{c}{$10^{11}\,h^{-2}\,L_{B_j,\odot}$}
&\multicolumn{1}{c}{}
&\multicolumn{1}{c}{$10^{11}\,h^{-2}\,L_{B_j,\odot}$}
\\
\multicolumn{1}{c}{(1)} 
&\multicolumn{1}{c}{(2)}
&\multicolumn{1}{c}{(3)}
&\multicolumn{1}{c}{(4)}
&\multicolumn{1}{c}{(5)}
&\multicolumn{1}{c}{(6)}
&\multicolumn{1}{c}{(7)} 
\\
\hline
A85       &$004140.6-091833$&1.94& 462&   26.75,   26.93& 403&   20.55,   23.41\\
A119      &$005617.9-011528$&1.36& 280&   11.36,   14.99& 370&   13.78,   17.48\\
A193      &$012505.4+084157$&1.45& 200&    6.64,   10.29&   -&    -   ,    -   \\
A194      &$012552.6-012008$& .68& 150&    1.73,    2.35& 218&    2.50,    3.03\\
A229      &$013914.5-033803$&1.01&  35&    5.66,    7.41&  50&    6.80,    8.15\\
A256      &$014804.5-035429$&1.09&  56&    9.68,    7.42&  85&   10.00,    7.61\\
A262      &$015246.7+360856$&1.05&1307&   15.12,   14.12&   -&    -   ,    -   \\
A295      &$020211.8-010603$& .72&   -&    -   ,    -   & 137&    5.94,    5.81\\
A400      &$025740.7+060048$&1.20& 351&    6.03,    6.61&   -&    -   ,    -   \\
A420      &$030916.8-113226$& .72&  25&    1.33,    1.66&  45&    2.73,    2.41\\
A458      &$034605.4-242040$&1.47&   -&    -   ,    -   & 114&   12.44,   17.82\\
A496      &$043331.8-131703$&1.37& 323&    5.38,    8.81& 526&    7.68,   11.09\\
A514      &$044830.1-203330$&1.76& 214&   33.19,   33.26& 294&   11.72,   16.10\\
A524      &$045743.2-194345$& .50&  27&    3.50,    1.73&  33&    2.20,    1.19\\
A978      &$102028.5-063050$&1.07& 165&    8.60,    9.72& 161&    5.51,    7.18\\
A999      &$102325.4+124958$& .56&  61&    3.71,    3.12&   -&    -   ,    -   \\
A1060     &$103631.2-272935$&1.22&   -&    -   ,    -   &3678&    8.80,   11.75\\
A1069     &$103937.0-083121$& .72&  67&    7.98,    5.60&  72&    6.25,    4.59\\
A1142     &$110154.8+101835$& .97& 133&    3.32,    4.43&   -&    -   ,    -   \\
A1146     &$110116.4-224413$&1.86&   -&    -   ,    -   & 148&   38.51,   43.15\\
A1185     &$111044.4+284145$&1.07& 380&    6.20,    7.68&   -&    -   ,    -   \\
A1228     &$112151.2+342201$& .34&  35&    2.17,    1.47&   -&    -   ,    -   \\
A1314     &$113428.0+490243$& .55&  97&    4.27,    4.56&   -&    -   ,    -   \\
A1631     &$125258.2-152111$&1.40& 357&   18.95,   12.49& 511&   22.25,   14.21\\
A1644     &$125723.3-172448$&1.52& 328&   11.92,   13.15& 524&   17.91,   17.97\\
A1656     &$125937.2+275712$&1.64&1136&   21.36,   26.80&   -&    -   ,    -   \\
A1795     &$134849.3+263347$&1.67& 239&   14.08,   19.52&   -&    -   ,    -   \\
A1809     &$135256.0+050760$&1.53& 144&   11.09,   18.77&   -&    -   ,    -   \\
A1983     &$145258.1+164129$& .99& 233&    5.97,    7.44&   -&    -   ,    -   \\
A1991     &$145432.6+183735$&1.26& 158&    7.55,   10.84&   -&    -   ,    -   \\
\hline
\end{tabular}
 
\end{center}
%\vspace{-2mm}
%\input{comm_tab3}
\end{minipage}
\begin{multicols}{2}

\end{multicols}
%%TAB3%\vspace{-6mm}
\hspace{-13mm}
\begin{minipage}{20cm}
\renewcommand{\arraystretch}{1.2}
\renewcommand{\tabcolsep}{1.2mm}
\begin{center}%\vspace{-3mm}
TABLE 3\\%\vspace{2mm}
{\sc Continued \\}
\footnotesize 
%%%%%%%%%%%%%%%%%%%%%%%%%%%%%%%%%%%%%%%%%%%%%%%%%%%%%%%%%%%%%%%%%%%%%%%
%
%                        TABLE 2
%
%%%%%%%%%%%%%%%%%%%%%%%%%%%%%%%%%%%%%%%%%%%%%%%%%%%%%%%%%%%%%%%%%%%%%%%
%
%                        TABLE 3
%
%%%%%%%%%%%%%%%%%%%%%%%%%%%%%%%%%%%%%%%%%%%%%%%%%%%%%%%%%%%%%%%%%%%%%%%
\begin{tabular}{lcrrcrc}
\hline \hline
\multicolumn{1}{c}{Name}
&\multicolumn{1}{c}{Center}
&\multicolumn{1}{c}{$R_{vir}$}
&\multicolumn{1}{c}{$N_{APS}$}
&\multicolumn{1}{c}{$L_{B_j,APS}$\tablenotemark{a}}
&\multicolumn{1}{c}{$N_{COSMOS}$}
&\multicolumn{1}{c}{$L_{B_j,COSMOS}$\tablenotemark{a,b}}
\\
\multicolumn{1}{c}{}
&\multicolumn{1}{c}{$\alpha(2000)$-$\delta(2000)$}
&\multicolumn{1}{c}{$h^{-1}\,Mpc$}
&\multicolumn{1}{c}{}
&\multicolumn{1}{c}{$10^{11}\,h^{-2}\,L_{B_j,\odot}$}
&\multicolumn{1}{c}{}
&\multicolumn{1}{c}{$10^{11}\,h^{-2}\,L_{B_j,\odot}$}
\\
\multicolumn{1}{c}{(1)} 
&\multicolumn{1}{c}{(2)}
&\multicolumn{1}{c}{(3)}
&\multicolumn{1}{c}{(4)}
&\multicolumn{1}{c}{(5)}
&\multicolumn{1}{c}{(6)}
&\multicolumn{1}{c}{(7)} 
\\
\hline
A1991     &$145432.6+183735$&1.26& 158&    7.55,   10.84&   -&    -   ,    -   \\
A2029     &$151056.7+054500$&2.33& 504&   49.67,   65.10&   -&    -   ,    -   \\
A2040     &$151246.4+072517$& .92& 158&    8.01,    7.87&   -&    -   ,    -   \\
A2048     &$151516.9+042229$&1.33& 109&   15.05,   15.90&   -&    -   ,    -   \\
A2079     &$152745.1+285508$&1.34& 172&   12.12,   13.89&   -&    -   ,    -   \\
A2092     &$153323.2+310856$&1.07& 106&    7.01,    6.77&   -&    -   ,    -   \\
A2107     &$153941.1+214843$&1.24& 242&    7.43,    9.53&   -&    -   ,    -   \\
A2124     &$154447.2+360440$&1.76& 286&   31.60,   40.53&   -&    -   ,    -   \\
A2142     &$155819.8+271435$&2.26& 275&   29.36,   44.31&   -&    -   ,    -   \\
A2151     &$160510.5+174522$&1.50& 626&   15.44,   18.97&   -&    -   ,    -   \\
A2197     &$162946.7+405036$&1.22& 546&   11.48,   15.02&   -&    -   ,    -   \\
A2199     &$162843.0+393043$&1.60& 797&   19.75,   25.61&   -&    -   ,    -   \\
A2353     &$213422.3-013507$&1.19&  36&   32.84,   36.90&  47&    6.65,   11.41\\
A2362     &$213901.3-141911$& .66&  54&    3.62,    4.09&  55&    1.56,    2.30\\
A2401     &$215821.1-200557$& .79& 165&   11.68,   11.38& 132&    5.65,    6.26\\
A2426     &$221411.9-101103$& .66&  71&   12.95,    4.01&  62&    6.77,    2.36\\
A2500     &$225351.8-253103$& .95&   -&    -   ,    -   &  63&    3.90,    2.64\\
A2554     &$231212.7-213050$&1.68&   -&    -   ,    -   & 181&   24.52,   26.66\\
A2569     &$231757.8-124635$& .98&   -&    -   ,    -   &  79&    6.34,    8.71\\
A2589     &$232401.4+164834$& .94& 106&    2.90,    4.18&   -&    -   ,    -   \\
A2634     &$233828.4+270133$&1.40& 507&   20.90,   24.40&   -&    -   ,    -   \\
A2644     &$234035.7-000350$& .36&   -&    -   ,    -   &  15&     .68,     .48\\
A2670     &$235414.6-102505$&1.70& 201&   14.42,   18.53& 245&   12.33,   17.24\\
A2715     &$000245.2-343932$& .93&   -&    -   ,    -   &  46&    6.25,    4.20\\
A2717     &$000310.8-355557$&1.08&   -&    -   ,    -   &  172&    3.1,    4.78\\
A2721     &$000607.9-344302$&1.61&   -&    -   ,    -   & 157&   19.71,   27.65\\
A2734     &$001126.6-285018$&1.26&   -&    -   ,    -   & 184&    4.64,    8.12\\
A2755     &$001736.9-351059$&1.54&   -&    -   ,    -   & 197&   15.88,   15.36\\
A2798     &$003733.1-283205$&1.42&  37&   10.87,    6.10& 108&   20.84,   10.48\\
A2799     &$003723.1-390750$& .84&   -&    -   ,    -   &  95&    3.80,    5.37\\
A2800     &$003804.2-250610$& .81&  42&    2.55,    3.42&  70&    3.53,    4.29\\
A2877     &$010948.3-455702$&1.77&   -&    -   ,    -   &1379&   10.94,   16.55\\
A2911     &$012604.0-375609$&1.09&   -&    -   ,    -   & 106&   11.47,    6.02\\
A3093     &$031057.5-472426$& .88&   -&    -   ,    -   &  71&    4.62,    5.28\\
A3094     &$031140.7-265908$&1.31&   -&    -   ,    -   & 223&   10.55,   12.30\\
A3111     &$031738.6-454656$& .32&   -&    -   ,    -   &  30&    3.10,     .98\\
A3122     &$032210.7-411905$&1.55&   -&    -   ,    -   & 223&    4.63,    9.60\\
A3126     &$032833.0-554241$&2.11&   -&    -   ,    -   & 281&   10.98,   24.02\\
A3128     &$033052.6-523023$&1.58&   -&    -   ,    -   & 434&   19.26,   21.86\\
A3142     &$033644.0-394616$&1.47&   -&    -   ,    -   & 132&   12.55,    9.99\\
A3151     &$034034.4-284043$& .47&   -&    -   ,    -   &  57&    4.10,    1.56\\
A3158     &$034259.7-533759$&1.95&   -&    -   ,    -   & 669&   19.38,   26.03\\
A3194     &$035908.8-301044$&1.61&   -&    -   ,    -   & 203&   16.99,   26.23\\
A3223     &$040809.7-310248$&1.29&   -&    -   ,    -   & 263&    9.39,   10.83\\
A3266     &$043046.8-613408$&2.21&   -&    -   ,    -   & 780&   24.76,   28.87\\
A3334     &$051749.1-583325$&1.39&   -&    -   ,    -   &  96&    6.68,   11.86\\
A3354     &$053442.5-284046$& .72&   -&    -   ,    -   & 102&    8.23,    5.43\\
A3360     &$054007.3-432358$&1.67&   -&    -   ,    -   & 140&    9.05,   17.06\\
A3376     &$060215.5-395625$&1.38&   -&    -   ,    -   & 288&    5.45,    8.95\\
A3381     &$060957.0-333320$& .59&   -&    -   ,    -   & 126&    4.61,    2.01\\
A3391     &$062617.6-534143$&1.33&   -&    -   ,    -   & 278&   16.26,   16.35\\
A3395     &$062735.6-542629$&1.70&   -&    -   ,    -   & 496&   16.04,   15.87\\
A3528N    &$125426.7-290017$& .92& 229&   14.14,    8.85&   -&    -   ,    -   \\
\hline
\end{tabular}
 
\end{center}
\end{minipage}
\begin{multicols}{2}

\end{multicols}
%%TAB3%\vspace{-6mm}
\hspace{-13mm}
\begin{minipage}{20cm}
\renewcommand{\arraystretch}{1.2}
\renewcommand{\tabcolsep}{1.2mm}
\begin{center}%\vspace{-3mm}
TABLE 3\\%\vspace{2mm}
{\sc Continued \\}
\footnotesize 
%%%%%%%%%%%%%%%%%%%%%%%%%%%%%%%%%%%%%%%%%%%%%%%%%%%%%%%%%%%%%%%%%%%%%%%
%
%                        TABLE 2
%
%%%%%%%%%%%%%%%%%%%%%%%%%%%%%%%%%%%%%%%%%%%%%%%%%%%%%%%%%%%%%%%%%%%%%%%
%
%                        TABLE 3
%
%%%%%%%%%%%%%%%%%%%%%%%%%%%%%%%%%%%%%%%%%%%%%%%%%%%%%%%%%%%%%%%%%%%%%%%
\begin{tabular}{lcrrcrc}
\hline \hline
\multicolumn{1}{c}{Name}
&\multicolumn{1}{c}{Center}
&\multicolumn{1}{c}{$R_{vir}$}
&\multicolumn{1}{c}{$N_{APS}$}
&\multicolumn{1}{c}{$L_{B_j,APS}$\tablenotemark{a}}
&\multicolumn{1}{c}{$N_{COSMOS}$}
&\multicolumn{1}{c}{$L_{B_j,COSMOS}$\tablenotemark{a,b}}
\\
\multicolumn{1}{c}{}
&\multicolumn{1}{c}{$\alpha(2000)$-$\delta(2000)$}
&\multicolumn{1}{c}{$h^{-1}\,Mpc$}
&\multicolumn{1}{c}{}
&\multicolumn{1}{c}{$10^{11}\,h^{-2}\,L_{B_j,\odot}$}
&\multicolumn{1}{c}{}
&\multicolumn{1}{c}{$10^{11}\,h^{-2}\,L_{B_j,\odot}$}
\\
\multicolumn{1}{c}{(1)} 
&\multicolumn{1}{c}{(2)}
&\multicolumn{1}{c}{(3)}
&\multicolumn{1}{c}{(4)}
&\multicolumn{1}{c}{(5)}
&\multicolumn{1}{c}{(6)}
&\multicolumn{1}{c}{(7)} 
\\
\hline
A3532     &$125715.3-302105$&1.48&   -&    -   ,    -   & 463&   18.92,   17.32\\
A3556     &$132418.2-314220$&1.28&   -&    -   ,    -   & 409&   16.22,   15.02\\
A3558     &$132755.5-312921$&1.95&   -&    -   ,    -   &1151&   50.30,   55.51\\
A3559     &$133011.5-293400$& .91& 186&    9.10,    9.60& 215&    9.52,   10.32\\
A3571     &$134720.8-325210$&2.09&   -&    -   ,    -   &1746&   35.97,   45.65\\
A3574     &$134849.3-302735$& .98&   -&    -   ,    -   &1303&    8.56,    9.25\\
A3651     &$195226.3-550815$&1.25&   -&    -   ,    -   & 348&   21.34,   19.65\\
A3667     &$201226.5-564840$&1.94&   -&    -   ,    -   & 882&   40.78,   40.10\\
A3693     &$203420.5-343311$& .96&   -&    -   ,    -   &  89&    8.31,    4.59\\
A3695     &$203445.0-354809$&1.56&   -&    -   ,    -   & 214&   20.65,   26.62\\
A3705     &$204210.2-351215$&1.75&   -&    -   ,    -   & 327&   29.61,   32.44\\
A3733     &$210135.2-280232$&1.22& 342&   10.34,   12.89& 301&    5.30,    8.27\\
A3744     &$210723.8-252558$&1.02&   -&    -   ,    -   & 340&    6.55,    7.76\\
A3809     &$214715.8-435523$& .96&   -&    -   ,    -   & 162&    5.78,    7.14\\
A3822     &$215414.8-575103$&1.62&   -&    -   ,    -   & 545&   36.26,   33.83\\
A3825     &$215819.6-601828$&1.40&   -&    -   ,    -   & 298&   16.96,   16.56\\
A3879     &$222759.3-685547$& .80&   -&    -   ,    -   &  81&    4.86,    5.15\\
A3880     &$222751.7-303419$&1.65&   -&    -   ,    -   & 365&   14.66,   19.75\\
A3921     &$225003.3-642351$& .98&   -&    -   ,    -   & 117&   12.05,   13.37\\
A4008     &$233020.5-391538$& .85&   -&    -   ,    -   & 126&    4.71,    5.34\\
A4010     &$233132.1-363010$&1.25&   -&    -   ,    -   & 102&    8.67,   10.56\\
A4053     &$235442.6-274115$&1.23&   -&    -   ,    -   & 136&    6.16,    4.60\\
A4067     &$235856.9-603730$&1.00&   -&    -   ,    -   &  84&    8.11,    8.74\\
S84       &$004925.3-293051$& .66&   -&    -   ,    -   &  50&    6.81,    5.08\\
S373      &$033603.9-351343$& .62&   -&    -   ,    -   &4150&    1.28,    3.68\\
S463      &$042911.7-535013$&1.22&   -&    -   ,    -   & 416&   15.80,   16.22\\
S721      &$130604.4-373704$&1.38&   -&    -   ,    -   & 463&   11.36,   12.25\\
S753      &$140316.0-340318$&1.07&   -&    -   ,    -   &1653&    5.10,    5.65\\
S805      &$185246.3-631441$&1.08&   -&    -   ,    -   &2133&    6.28,    7.59\\
S987      &$220154.8-222433$&1.35&   -&    -   ,    -   & 206&   21.45,   25.42\\
S1157/C67 &$235139.8-342714$&1.16&   -&    -   ,    -   & 191&    8.87,    5.36\\
AWM4      &$160455.3+235627$& .24&  26&    1.62,     .75&   -&    -   ,    -   \\
CL2335-26 &$233753.1+271047$&1.20&  51&   58.34,   51.83&   -&    -   ,    -   \\
DC0003-50 &$000604.6-503842$& .70&   -&    -   ,    -   & 146&    3.17,    3.96\\
Eridanus  &$034014.6-183725$& .53&   -&    -   ,    -   &1536&     .52,     .96\\
MKW1      &$100044.2-025741$& .45&   -&    -   ,    -   & 185&    1.51,    1.82\\
MKW6A     &$141439.7+030753$& .55&  80&    1.38,    2.04&   -&    -   ,    -   \\   		   
S49-145   &$020734.9+020814$& .74&-  &     -  ,   -       &279&    3.47,     (3.68)\\
S49-142   &$032044.7-010215$& .24& 21&     .41,      (.41)& 16&     .21,     (.26) \\
N45-384   &$092751.8+295956$& .46&149&    1.57,        .90&-  &     -  ,     -     \\
N34-172   &$100032.1-025727$& .60&158&    1.05,       1.43&-&     -  ,     -     \\
N56-393   &$101352.0+384007$& .62&116&    1.50,     (1.87)&-&     -  ,     -     \\
N79-278   &$113754.4+215823$& .57& 63&    1.24,       1.60&-&     -  ,     -     \\
N67-312   &$114204.6+101820$& .26& 53&     .75,      (.71)&-&     -  ,     -     \\
N56-371   &$114503.5-013938$& .64&162&    1.61,     (1.95)&246&    1.90,   (2.21)  \\
N79-280   &$114618.5+330919$& .17& 25&    1.07,      (.89)&-&     -  ,     -     \\
N56-392   &$114938.9-033135$& .34& 31&     .73,        .84& 65&    1.42,    1.36   \\
N79-298   &$115752.3+251018$& .30& 71&     .35,      (.42)&-&     -  ,     -     \\
N79-299B  &$120409.5+201318$& .73&234&    3.13,     3.00  &-&     -  ,     -     \\
N67-335   &$120421.7+015019$&1.03&490&   1.41,     1.60  &-&    -  ,     -     \\
N79-299A  &$120551.2+203219$& .78&200&    2.77,     2.86  &-&     -  ,     -     \\
\hline
\end{tabular}
 
\end{center}
\end{minipage}
\begin{multicols}{2}

\end{multicols}
%%TAB3%\vspace{-6mm}
\hspace{-13mm}
\begin{minipage}{20cm}
\renewcommand{\arraystretch}{1.2}
\renewcommand{\tabcolsep}{1.2mm}
\begin{center}%\vspace{-3mm}
TABLE 3\\%\vspace{2mm}
{\sc Continued \\}
\footnotesize 
%%%%%%%%%%%%%%%%%%%%%%%%%%%%%%%%%%%%%%%%%%%%%%%%%%%%%%%%%%%%%%%%%%%%%%%
%
%                        TABLE 2
%
%%%%%%%%%%%%%%%%%%%%%%%%%%%%%%%%%%%%%%%%%%%%%%%%%%%%%%%%%%%%%%%%%%%%%%%
%
%                        TABLE 3
%
%%%%%%%%%%%%%%%%%%%%%%%%%%%%%%%%%%%%%%%%%%%%%%%%%%%%%%%%%%%%%%%%%%%%%%%
\begin{tabular}{lcrrcrc}
\hline \hline
\multicolumn{1}{c}{Name}
&\multicolumn{1}{c}{Center}
&\multicolumn{1}{c}{$R_{vir}$}
&\multicolumn{1}{c}{$N_{APS}$}
&\multicolumn{1}{c}{$L_{B_j,APS}$\tablenotemark{a}}
&\multicolumn{1}{c}{$N_{COSMOS}$}
&\multicolumn{1}{c}{$L_{B_j,COSMOS}$\tablenotemark{a,b}}
\\
\multicolumn{1}{c}{}
&\multicolumn{1}{c}{$\alpha(2000)$-$\delta(2000)$}
&\multicolumn{1}{c}{$h^{-1}\,Mpc$}
&\multicolumn{1}{c}{}
&\multicolumn{1}{c}{$10^{11}\,h^{-2}\,L_{B_j,\odot}$}
&\multicolumn{1}{c}{}
&\multicolumn{1}{c}{$10^{11}\,h^{-2}\,L_{B_j,\odot}$}
\\
\multicolumn{1}{c}{(1)} 
&\multicolumn{1}{c}{(2)}
&\multicolumn{1}{c}{(3)}
&\multicolumn{1}{c}{(4)}
&\multicolumn{1}{c}{(5)}
&\multicolumn{1}{c}{(6)}
&\multicolumn{1}{c}{(7)} 
\\
\hline
N79-283   &$121954.8+282521$& .72&183&    6.99,     2.09  &-&     -  ,     -     \\
N79-292   &$122414.7+092024$& .64& 80&     .36,      .43  &-&     -  ,     -     \\
N67-333   &$130425.3+075454$& .98&110&    2.63,     2.49  &-&     -  ,     -     \\
N67-323   &$130526.5+533356$& .16& 11&     .48,      (.42)&-&     -  ,     -     \\
N67-317   &$131349.0+065709$& .45& 64&     .42,      (.63)&-&     -  ,     -     \\
N79-270   &$131719.3+203711$& .26& 26&     .41,      (.43)&-&     -  ,     -     \\
N79-296   &$132922.3+114731$& .78&289&    2.72,     2.08  &-&     -  ,     -     \\
N67-329   &$133236.4+072036$& .35& 39&     .76,      (.81)&-&     -  ,     -     \\
N79-297   &$135524.7+250320$& .41& 65&    1.75,     (1.67)&-&     -  ,     -     \\
N79-276   &$135622.3+283123$&1.08&206&    1.92,     1.78  &-&    -  ,     -     \\
N67-336   &$140304.0+092635$& .54&159&    2.99,     2.91  &-&     -  ,     -     \\
N67-325   &$140958.5+173251$& .53&199&    1.76,     (2.00)&-&     -  ,     -     \\
N67-326   &$142814.1+255038$& .51&199&     .67,      (.97)&-&     -  ,     -     \\
N67-309   &$142831.6+112238$& .68&119&    1.68,     (2.08)&-&     -  ,     -     \\
N56-394   &$143400.9+034453$& .59&109&    1.41,     1.73  &-&     -  ,     -     \\
N56-395   &$144043.2+032712$& .95&302&    1.96,     3.06  &-&     -  ,     -     \\
N56-381   &$144700.4+113529$& .52& 62&     .31,      (.64)&-&     -  ,     -     \\
N45-381   &$151311.6+042850$& .30& 41&    1.60,     (1.40)&-&     -  ,     -     \\
N45-363   &$155746.9+161627$& .90&178&    4.52,     2.67  &-&     -  ,     -     \\
N45-389   &$161739.2+350545$&1.31&408&    6.46,     9.40  &-&     -  ,     -     \\
N34-171   &$164135.4+575021$& .43& 97&     .35,      (.57)&-&     -  ,     -     \\
N34-175   &$171521.4+572243$&1.10&976&   12.18,    (12.11)&-&     -  ,     -     \\
NGC 533   &$012529.9+014537$& .93&311&    1.33,     2.71  &597&    1.65,    2.83   \\
HCG 42    &$100018.6-193860$& .42&-  &     -  ,        -  &554&    2.29,    2.40   \\
NGC 4325  &$122304.6+103352$& .53& 57&     .44,      .76  &-&     -  ,     -     \\
HCG 62    &$125305.1-091247$& .79&449&    2.04,     2.79  &978&    5.03,    5.59   \\
NGC 5129  &$132425.2+135455$& .59&129&    1.79,     2.27  &-&     -  ,     -     \\
NGC 491   &$012102.7-340358$& .18&-  &     -  ,        -  & 51&     .34,     (.32) \\
NGC 664   &$014314.4+041319$& .30& 27&     .23,     (.32) &-&     -  ,     -     \\
\hline
\end{tabular}
 
\end{center}
\vspace{-2mm}

{\footnotesize\parindent=3mm 
$^a$~Both alternative estimates of luminosity
are given ($L_{Bj,c}$, $L_{Bj,f}$).
Values between brackets 
are $L_{Bj,f}$, which are not
based on individual member fraction estimates, but on median values.\\
$^b$~We report values computed by G00 for clusters (C-CL 
sample), too.}

\end{minipage}
\begin{multicols}{2}

Figure~5 compares the luminosities derived from the COSMOS catalog
with those derived from the APS catalog for the 27 systems having both
photometries available.  For both the two alternative estimates
(panels a and b) the fitted straight lines in the logarithmic plane
are consistent with the one--to--one relation within $1\sigma$ (using
the unweighted bisecting fit; cf. Isobe et al. 1990).  This fair
agreement supports the homogeneity of our luminosity estimates,
although recovered from two different photometric catalogs, and
justify their combination.  When both APS and COSMOS luminosities are
available, we consider their average.

Figure~6 compares the two alternative luminosity estimates $L_{Bj,c}$
and $L_{Bj,f}$ for all 119 CLs and those 22 PSs for which we compute
individual member fractions.  The (bisecting) fit is consistent with
the one--to--one relation, within $2\sigma$, indicating that any
systematic bias connected to the fore/background correction should not
seriously pollute our analysis.  More particularly, Figure~6 shows no
systematic overestimate of $L_{B_j,c}$ with respect to $L_{B_j,f}$, as
expected when subtracting the mean field luminosity density from
cluster regions, when clusters are selected as overdensities in a
projected galaxy distribution. This is due to two main reasons.
First, systematic foreground/background contamination for nearby
clusters is small (e.g. only $\sim10\%$ for ENACS, Katgert et
al. 1996); second, in this study we exclude, a priori, clusters
showing two significant peaks in the velocity distribution, which are
the most contaminated clusters (cf. \SS~2.1 and 3.1.1).  The
comparison in Figure~6 suggests that the majority of clusters has
$L_{B_j,f}>L_{B_j,c}$, although for a small amount
($L_{B_j,f}/L_{B_j,c}=1.1$, median value), probably because that the
member fractions we determine in the redshift samples are slightly
larger than the appropriate ones for the (deeper) magnitude
samples. The distribution of the scatter is not symmetric with respect
the one--to--one relation, thus suggesting competition between two
different sources of errors. This supports the use, in our final
results, of the average of the two estimates.

As for the error estimates, the scatter in the $L_{Bj,c}-L_{Bj,f}$
relation ($\sim 25\%$) provides an estimate of the random errors due
to the variations of local field, i.e. to the procedure adopted, while
the scatter in the COSMOS--APS comparison ($\sim 40\%$ and $\sim 30\%$
in the two cases) also gives an idea of the random errors connected
with the photometry of the catalogs.  By adding in the quadrature the
two sources of errors, we assume that the error on each individual
luminosity is $\lesssim 50\%$.

%\end{multicols}
%\begin{figure}
\includegraphics{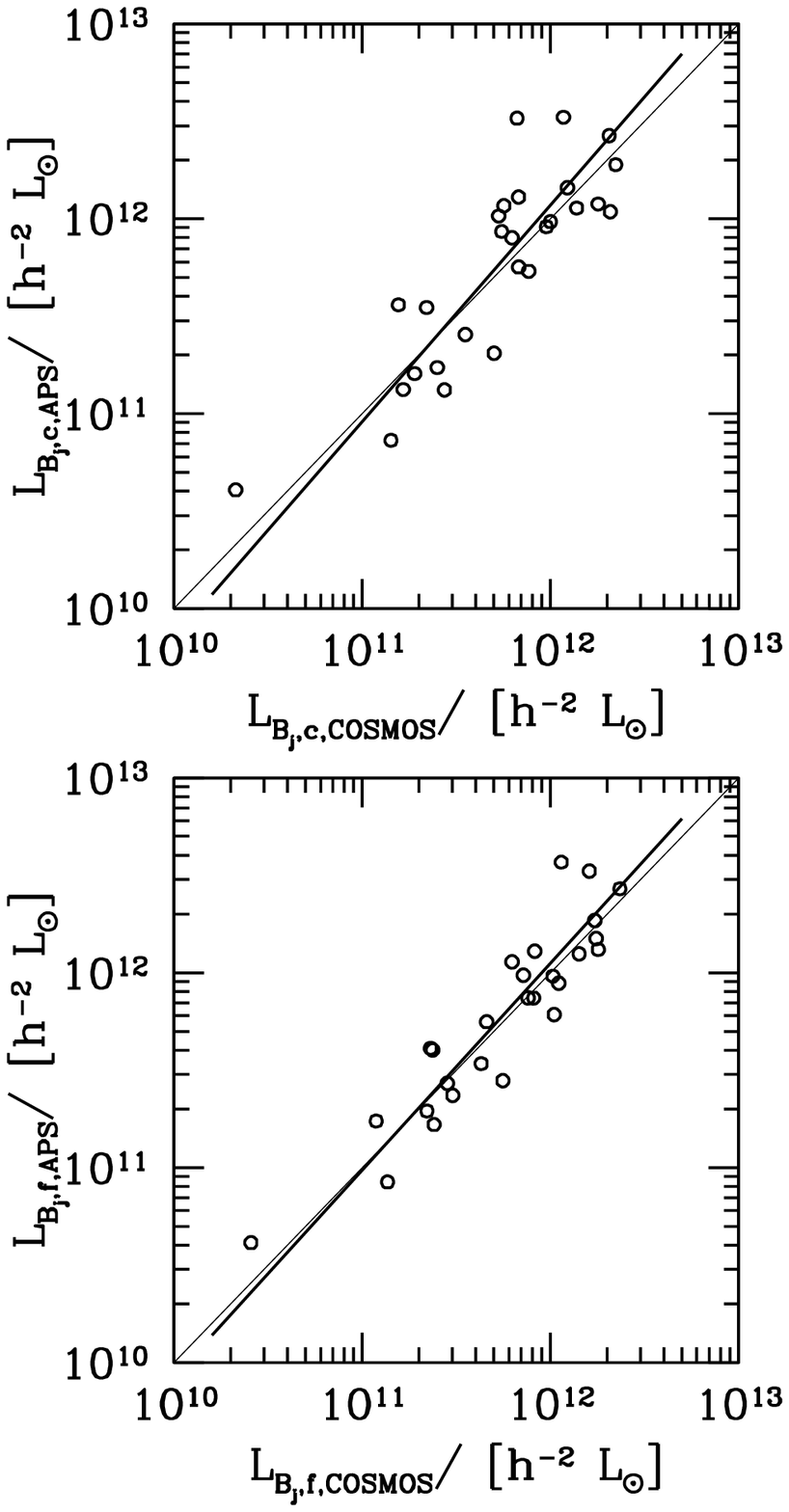}
$\ \ \ \ \ \ $\\
\vspace{15truecm}
$\ \ \ $\\
%\vspace{1cm}
{\small\parindent=3.5mm {Fig.}~5.---
Comparison between the luminosities derived from
the APS catalog with those derived from the COSMOS catalog for the 27 galaxy
systems with both photometries available, and for both the two
alternative estimates of the fore/background correction: by using the
mean counts (top--panel) or the member fraction (bottom--panel).
The solid lines represent the (bisecting) fitting to be
compared to the one--to--one relation (the faint line).
}
\vspace{5mm}
%\begin{multicols}{2}

\subsection{Mass and Luminosity for NOG Groups}

Here we compute mass and luminosity for each group of
NOG group catalogs, PG and HG.

The observational determination of group $M/L$ encounters several
additional problems.  These problems arise in the estimate of mass and
are mainly due to the poor number of group members and to the
uncertainties in the dynamical stage.  In fact, although group cores
are close to virialization or virialized (ZM98; Zabludoff \& Mulchaey
1998b), the sampling area of groups identified in three--dimensional
galaxy catalogs is well outside their likely virialized region (e.g.,
Girardi \& Giuricin 2000, hereafter GG00; Carlberg et al. 2001a).
Also older works indicated that, by considering their whole sampled
region, these groups cannot be considered virialized systems, but
rather described as being in a phase of collapse (e.g., Giuricin et
al. 1988; Diaferio, Geller, \& Ferrari 1993; Mamon 1994).  The small
number of galaxies prevents us from applying refined analyses such as
those used for clusters and well--sampled groups, e.g. the member
selection or the analysis of velocity dispersion profiles (cf. G98 and
reference therein; ZM98; Mahdavi et al. 1999).  Above all, the small
number of galaxies prevent us from working in smaller, quasi
virialized, group regions.

%\end{multicols}
%\begin{figure}
\includegraphics{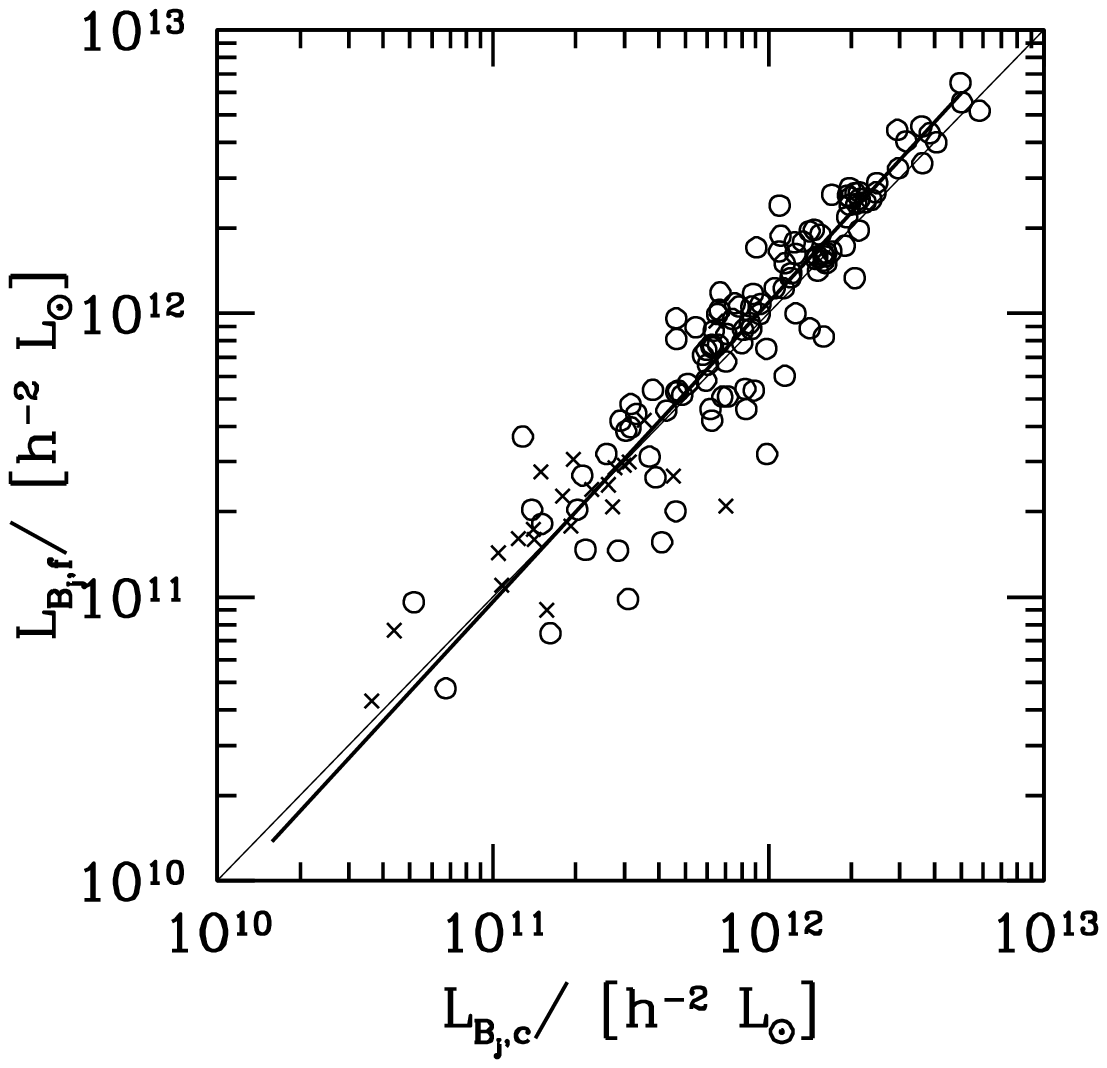}
$\ \ \ \ \ \ $\\
\vspace{8truecm}
$\ \ \ $\\
%\vspace{1cm}
{\small\parindent=3.5mm {Fig.}~6.---
Comparison between the two
alternative luminosity estimates based on the two different
fore/background corrections: by using the
mean counts ($L_{B_J,c}$) or the member fraction ($L_{B_J,f}$).
Circles and crosses indicate clusters and poor systems, respectively.
The solid line represents the (bisecting) fitting to be
compared to the one--to--one relation (the faint line).
}
\vspace{5mm}
%\begin{multicols}{2}

Therefore, for computing group masses we could not apply the procedure
used in \S~3.1, but we rather use the procedure recently adopted by
GG00.  Here we summarize the main steps of this procedure.

First of all, we do not perform any procedure of member selection, but
we rely on the group membership as assigned by Giuricin et al. (2000).
The comparison of the results coming from the two catalogs of groups
identified in the same galaxy catalog, but with two different member
assignments, will allow us to check a posteriori the effect of the
membership procedure.

Then we compute the l.o.s. velocity dispersion, $\sigma_v$, and the
(projected) radius used in the virial theorem, $R_{PV}$, in the same way
as performed in \S~3.1; in particular, for each galaxy we assume a
typical velocity error of 30 \kss. The corresponding virial mass is
$M_V=3\pi \sigma_v^2 R_{PV}/(2G)$.  When the correction for velocity
errors leads to a negative value of $\sigma_v$, $\sigma_v$ and mass
are considered null.

To take into account the dynamical state of groups we use the method
proposed by Giuricin et al. (1988). This method is based on the
classical model of spherical collapse where the initial density
fluctuation grows, lagging behind the cosmic expansion when it breaks
away from the Hubble flow, and begins to collapse, and  the authors used
the results of very simple numerical simulations of galaxy systems
(Giuricin et al. 1984; the limits of this model are discussed in \S
7.1 of GG00).  Using the above method, the value of $A$, which is
needed to recover corrected masses as $M_{VC}=(1/2A)M_V$, can be
inferred from the estimate of the presently observed crossing
time.  In particular, the precise value of $A$ depends on the
background cosmology: for our reference model we obtain a typical
correction of $0\%$ and $30\%$ for PGs and HGs, respectively (cf. GG00
for other examples).

As for the luminosity computation, we gain a great advantage with
respect to poor systems in \S~3.2. In fact, the NOG catalog is already
homogeneized for photometry; the total blue magnitudes, $B$, are
already corrected (for Galactic absorption, K--dimming, and internal
absorption); and the known membership avoids the problem of
fore/background contamination. Taking into account these differences,
we apply a procedure similar to that outlined in \S~3.2.3, i.e.: we
compute observed group luminosities, $L_{B,obs}$, (assuming
$B_{\odot}=5.48$), which here refer to the whole sampled region; we
correct for $97\%$ NOG incompleteness, obtaining $L_{B,compl}$; we
include the luminosity of faint galaxies below the magnitude completeness
limit to obtain total luminosity $L_{B,tot}$, with $M_{B}^*=-19.97$
for the characteristic magnitude and $\alpha=-1.16$ for the slope
(Giuricin et al. 2000). The median correction for faint galaxies is
less than $1\%$.

For the two NOG catalogs, PG and HG, as well as some subsamples: PG5
and HG5, PG7 and HG7, i.e. groups having at least 5 and 7 members
respectively, we give median values (and $90\%$ c.l. error bands) for
interesting physical quantities.  Table~4 lists the sample name
[Col.~(1)]; $N_G$, the number of groups [Col.~(2)], and the median
values for: $N_m$, the number of member galaxies [Col.~(2)];
$R_{max}$, the group size which is the projected distance of the most
distant galaxy from the group center, here computed as the biweight
center [Col.~(4)]; $\sigma_v$, the l.o.s. velocity dispersion
[Col.~(5)]; $R_{PV}$, the projected radius used in the virial theorem
[Col.~(6)]; $M_V$ and $M_{VC}$, the virial mass, before and after the
correction for the dynamical status [Cols.~(7) and (8)]; $L_{B,tot}$,
the blue luminosity [Cols.~(9)]. Both mass and luminosity refer to the
whole sampled region.

\end{multicols}
%%TAB4%\vspace{-6mm}
\hspace{-13mm}
\begin{minipage}{20cm}
\renewcommand{\arraystretch}{1.2}
\renewcommand{\tabcolsep}{1.2mm}
\begin{center}%\vspace{-3mm}
TABLE 4\\%\vspace{2mm}
{\sc Results for NOG Groups \\}
\footnotesize 

%%%%%%%%%%%%%%%%%%%%%%%%%%%%%%%%%%%%%%%%%%%%%%%%%%%%%%%%%%%%%%%%%%%%%%%
%
%                        TABLE 4
%
%%%%%%%%%%%%%%%%%%%%%%%%%%%%%%%%%%%%%%%%%%%%%%%%%%%%%%%%%%%%%%%%%%%%%%%
\begin{tabular}{lrrrrrrrr}
\hline \hline
\multicolumn{1}{c}{Cat.}
&\multicolumn{1}{c}{$N_G$}
&\multicolumn{1}{c}{$N_m$}
&\multicolumn{1}{c}{$R_{max}$}
&\multicolumn{1}{c}{$\sigma_v$}
&\multicolumn{1}{c}{$R_{PV}$}
&\multicolumn{1}{c}{$M_V$}
&\multicolumn{1}{c}{$M_{CV}$}
&\multicolumn{1}{c}{$L_{B,tot}$}
\\
\multicolumn{1}{c}{}
&\multicolumn{1}{c}{}
&\multicolumn{1}{c}{}
&\multicolumn{1}{c}{$h^{-1}\,Mpc$}
&\multicolumn{1}{c}{$km\,s^{-1}$}
&\multicolumn{1}{c}{$h^{-1}\,Mpc$}
&\multicolumn{1}{c}{$10^{13}\,h^{-1}\,M_{\odot}$}
&\multicolumn{1}{c}{$10^{13}\,h^{-1}\,M_{\odot}$}
&\multicolumn{1}{c}{$10^{11}\,h^{-2}\,L_{B,\odot}$}
\\
\multicolumn{1}{c}{(1)} 
&\multicolumn{1}{c}{(2)}
&\multicolumn{1}{c}{(3)}
&\multicolumn{1}{c}{(4)}
&\multicolumn{1}{c}{(5)}
&\multicolumn{1}{c}{(6)}
&\multicolumn{1}{c}{(7)}
&\multicolumn{1}{c}{(8)}
&\multicolumn{1}{c}{(9)} 
\\
\hline
PG &513&4&$0.46^{+0.04}_{-0.04}$&$132^{+12}_{-13}$&$0.49^{+0.03}_{-0.05}$&$0.81^{+0.17}_{-0.15}$&$0.85^{+0.22}_{-0.17}$&$0.52^{+0.04}_{-0.03}$\\
PG5&208&7&$0.60^{+0.06}_{-0.03}$&$162^{+17}_{-22}$&$0.53^{+0.04}_{-0.04}$&$1.49^{+0.33}_{-0.43}$&$1.47^{+0.34}_{-0.23}$&$0.82^{+0.08}_{-0.09}$\\
PG7&112&10&$0.76^{+0.04}_{-0.09}$&$199^{+11}_{-19}$&$0.56^{+0.08}_{-0.04}$&$2.44^{+0.73}_{-0.62}$&$2.57^{+0.58}_{-0.68}$&$1.15^{+0.14}_{-0.21}$\\
HG &475&4&$0.58^{+0.04}_{-0.03}$&$ 84^{+7}_{-6}$&$0.60^{+0.05}_{-0.05}$&$0.43^{+0.07}_{-0.12}$&$0.53^{+0.16}_{-0.09}$&$0.52^{+0.04}_{-0.06}$\\
HG5&190&7&$0.81^{+0.06}_{-0.07}$&$106^{+8}_{-7}$&$0.66^{+0.05}_{-0.06}$&$0.78^{+0.22}_{-0.12}$&$1.05^{+0.17}_{-0.27}$&$0.92^{+0.16}_{-0.08}$\\
HG7&103&10&$0.98^{+0.13}_{-0.13}$&$120^{+11}_{-8}$&$0.71^{+0.08}_{-0.08}$&$1.22^{+0.33}_{-0.26}$&$1.57^{+0.48}_{-0.46}$&$1.23^{+0.28}_{-0.13}$\\
\hline
\end{tabular}
 
\end{center}
\end{minipage}
\begin{multicols}{2}

Typical mass errors are very large and vary with group mass: e.g., we
find a median mass error of $\sim130\%$ on the whole catalogs, and
only $\sim70\%$ for groups with at least five members.  Both mass and
luminosity could have large systematic errors connected to the
algorithm of identification: for instance the typical mass of PG and
HG differs for a $60\%$, while the luminosity seems enough robust.

\subsection{NOG Groups vs Other Galaxy Systems}

As for a reliable comparison with clusters analyzed by G00, and other
galaxy systems analyzed in \S~3.1 and 3.2, we should rescale both
group mass and luminosity to the region within $R_{vir}$. The paucity
of data does not allow us to make an individual correction: we apply a
mean correction to all groups by using the procedure outlined by G00.

PGs and HGs are identified by using a number density contrast $(\delta
\rho/\rho)_g=80$ and a luminosity density contrast $(\delta
\rho/\rho)_g=45$, respectively, which are comparable to the matter
density contrast, $\delta \rho/\rho$, since the biasing factor $b=
(\delta \rho/ \rho)_g/(\delta\rho/ \rho)$ is roughly one (e.g., from
$b=1/\sigma_8$ and $\sigma_8$ value from Eke et al. 1996; Girardi et
al. 1998a).  These values of $\delta \rho/ \rho$ for $PG$ and $HG$
groups are much smaller than the values of $\sim 350$ expected within
the virialized region (e.g., Eke et al. 1996).

After assuming that groups have a common radial profile (Fasano et
al. 1993), we can roughly estimate the number fraction of members
contained in the virialized region.  For each of the two catalogs,
Figure~7 plots the cumulative distributions of the projected galaxy
distances from the group center, combining together data of all
groups. To combine the galaxies of all groups we divide each galaxy
distance by the projected radius, $R_{PV}$, of its group and then we
normalize to the mean $<R_{PV}/R_{max}>$ of the catalog.  From
Figure~7 one can infer the fraction of the number of galaxies,
i.e. the fraction of group luminosity and mass if galaxy number
distribution traces luminosity and mass, contained within each radius.
We are interested in determining the radius, and the corresponding
number fraction, for which one obtains a density enhancement which is
large enough to reach the density contrast expected in the virialized
region.  In the case of PG, the virialization density is obtained
within a radius smaller by $\sim 55\%$, which contains $\sim 73\%$
fewer galaxies; in fact the density in these central regions is
$0.73/(0.55^3)\times80=\sim4.4\times 80\sim 350$.  In the case of HG,
similar arguments show that the virialization density is obtained
within a radius smaller by $\sim 42\%$, which contains $\sim 57\%$
fewer galaxies. The direct application of $R_{vir}$ definition (eq.~1)
would give similar small virialization radii: e.g., for PG groups the
median value of $\sigma_v\sim132$ \ks corresponds to a value of
$R_{vir}=0.26$ \h to be compared with the sampling area of groups
$R_{max}=0.46$ \hh, i.e. $\sim 56\%$ smaller (cf. also Carlberg et
al. 2001a for CNOC2 groups).

%\end{multicols}
%\begin{figure}
\includegraphics{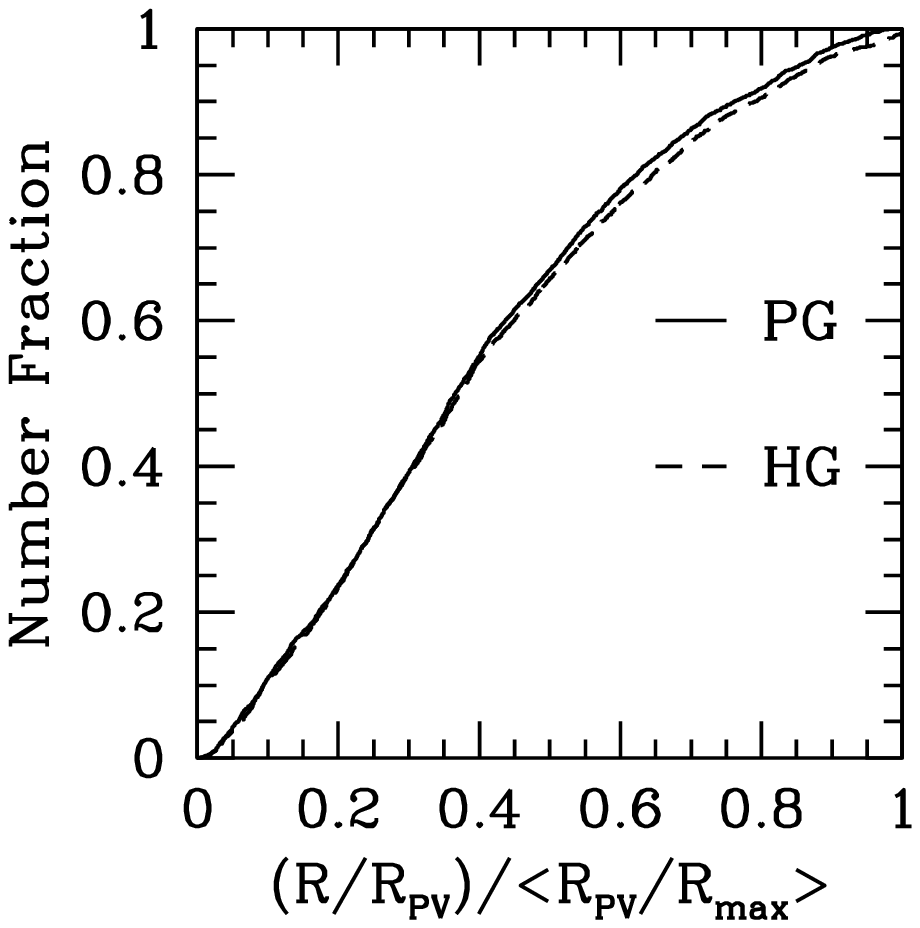}
$\ \ \ \ \ \ $\\
\vspace{8truecm}
$\ \ \ $\\
%\vspace{1cm}
{\small\parindent=3.5mm {Fig.}~7.---
For both PG and HG catalogs we show the cumulative
distribution of the projected galaxy distances from the group
center. To combine the galaxies of all groups we divide each galaxy
distance by the projected radius, $R_{PV}$, of its group.  Moreover,
we also normalize the distances to the median $<R_{PV}/R_{max}>$ of
the catalog.
}
\vspace{5mm}
%\begin{multicols}{2}

Owing to all uncertainties involved, we consider a similar
intermediate correction for all groups of both catalogs and assume
these fiducial values for virialized regions: $R_{vir}=0.5\times
R_{max}$, $M=0.65\times M_{VC}$, and $L_{B}=0.65\times L_{B,tot}$.

Moreover, when comparing results in the $B$ band with those in the
$B_j$ band, we assume that
$(L_{Bj}/L_{Bj,\odot})/(L_{B}/L_{B,\odot})=1.1$ in agreement with the
relation $B_j-B=0.28(B-V)$ of Blair \& Gilmore 1982 (see also
Metcalfe, Fong, \& Shanks 1995; Maddox, Efstathiou, \& Sutherland
1990), where we take $(B_j-B)_{\odot}=0.15$ and a mean value of
$(B-V)=0.9$ mag for cluster members.

Finally, we attempt a comparison of the results for those systems
catalogued both as NOG groups and as clusters or poor systems (CL+PS).
We identify 13 groups for PG and 12 for HG: A194/PG72/HG78,
A262/PG102/HG113, A3574/PG753/HG725,\\ Eridanus/PG202/HG207,
S373/PG201/HG208, S753/PG770/HG752, S805/PG929/HG898,
N67-312/PG581/HG554, N79-298/PG605/HG581,\\ N67-336/PG776/HG753,
N67-325/PG787/HG766, N67-326/PG809, N34-171/PG896,\\ HCG62/HG646.  Both
mass and luminosity of NOG groups correlate with the corresponding
CL+PS values, but there is a large scatter.  As for PGs, masses are
comparable and mass--to--light ratios are slightly larger:
$M_{PG}/M_{CL+PS}=0.91~(0.70-1.95)$ and\\
$(M/L)_{PG}/(M/L)_{CL+PS}=1.36~(0.78-1.71)$, median values with $90\%$
c.l. error bands.  As for HGs, masses are smaller and mass--to--light
ratios comparable: $M_{HG}/M_{CL+PS}=0.48~(0.35-0.63)$ and
$(M/L)_{HG}/(M/L)_{CL+PS}=0.87~(0.44-1.58)$. However, the sample is so
small and sparse that this comparison is not useful for understanding
which catalog is the more consistent with the treatment of other
galaxy systems.

\subsection{The Combined Sample}

Our comparison between groups and other systems does not enable us to
express a preference to one or the other of the two NOG catalogs
(cf. the above section).  Indeed, both algorithms present known
problems.  It has been suggested that the drawback of percolation
methods is the inclusion in the catalogs of possible non--physical
systems, like a long galaxy filament aligned close to the l.o.s.,
which could give large mass estimates, while the drawback of
hierarchical methods is the splitting of galaxy clusters into various
subunits, which give small mass estimates (e.g., Gourgoulhon et
al. 1992; Giuricin et al. 2000).  Therefore we choose to consider only
common groups, i.e.  those groups which are identified by both the
algorithms (cf. cross--identifications of Tables~5 and 7 of Giuricin
et al. 2000), averaging the corresponding estimates of physical
quantities. Avoiding PGs which are split into two or more HGs and,
viceversa, groups with null mass, as well as groups which are already
present in the CL+PS sample, we obtain 296 groups.

Moreover, the physical reality of the very poor detected groups has
often been discussed in the literature.  In particular, the efficiency
of the percolation algorithm has been repeatedly checked in the
literature, showing that an appreciable fraction of the poorer groups,
those with $N_m<5$ members, is false (i.e. unbound density
fluctuations), whereas the richer groups almost always correspond to
real systems (e.g., Ramella, Geller, \& Huchra 1989; Ramella et
al. 1995; Mahdavi et al. 1997; Nolthenius, Klypin, \& Primack 1997;
Diaferio et al. 1999).  Therefore, among the 296 common groups we
consider the 132 groups having more than five members (hereafter GROUP
sample). By combining GROUPs with other systems we obtain a fiducial
combined sample of 294 systems (CL+PS+GROUP).

\end{multicols}
%%TAB5%\vspace{-6mm}
\hspace{-13mm}
\begin{minipage}{20cm}
\renewcommand{\arraystretch}{1.2}
\renewcommand{\tabcolsep}{1.2mm}
\begin{center}%\vspace{-3mm}
TABLE 5\\%\vspace{2mm}
{\sc M/L Values for the Combined Sample \\}
\footnotesize 

%%%%%%%%%%%%%%%%%%%%%%%%%%%%%%%%%%%%%%%%%%%%%%%%%%%%%%%%%%%%%%%%%%%%%%%
%
%                        TABLE 5
%
%%%%%%%%%%%%%%%%%%%%%%%%%%%%%%%%%%%%%%%%%%%%%%%%%%%%%%%%%%%%%%%%%%%%%%%
\begin{tabular}{lclclc}
\hline \hline
\multicolumn{1}{c}{Name}
&\multicolumn{1}{c}{$lg\left(\frac{M}{M_{\odot}}\right),\frac{M/L_B}{(M/L_B)_{\odot}}$}
&\multicolumn{1}{c}{Name}
&\multicolumn{1}{c}{$lg\left(\frac{M}{M_{\odot}}\right),\frac{M/L_B}{(M/L_B)_{\odot}}$}
&\multicolumn{1}{c}{Name}
&\multicolumn{1}{c}{$lg\left(\frac{M}{M_{\odot}}\right),\frac{M/L_B}{(M/L_B)_{\odot}}$}
\\
\multicolumn{1}{c}{}
&\multicolumn{1}{c}{}
&\multicolumn{1}{c}{}
&\multicolumn{1}{c}{}
&\multicolumn{1}{c}{}
&\multicolumn{1}{c}{}
\\
\multicolumn{1}{c}{(1)} 
&\multicolumn{1}{c}{(2)}
&\multicolumn{1}{c}{(3)}
&\multicolumn{1}{c}{(4)} 
&\multicolumn{1}{c}{(5)}
&\multicolumn{1}{c}{(6)}
\\
\hline
A85       &     14.99,      445&A119      &     14.40,      191&A193      &     14.58,      490\\
A194      &     13.78,      274&A229      &     14.30,      315&A256      &     14.19,      198\\
A262      &     14.12,       99&A295      &     13.89,      145&A400      &     14.40,      433\\
A420      &     13.88,      413&A458      &     14.74,      401&A496      &     14.50,      427\\
A514      &     14.91,      376&A524      &     13.30,      103&A978      &     14.37,      332\\
A999      &     13.49,      100&A1060     &     14.28,      204&A1069     &     13.88,      136\\
A1142     &     14.29,      552&A1146     &     14.87,      199&A1185     &     14.02,      167\\
A1228     &     12.79,       37&A1314     &     13.49,       78&A1631     &     14.70,      328\\
A1644     &     14.65,      325&A1656     &     14.70,      227&A1795     &     14.77,      384\\
A1809     &     14.66,      338&A1983     &     14.22,      271&A1991     &     14.27,      222\\
A2029     &     14.83,      131&A2040     &     14.24,      243&A2048     &     14.40,      178\\
A2079     &     14.67,      391&A2092     &     14.12,      212&A2107     &     14.42,      340\\
A2124     &     14.84,      210&A2142     &     15.25,      533&A2151     &     14.75,      361\\
A2197     &     14.55,      296&A2199     &     14.76,      277&A2353     &     14.31,      103\\
A2362     &     13.79,      236&A2401     &     13.98,      121&A2426     &     13.68,       80\\
A2500     &     14.01,      340&A2554     &     14.78,      261&A2569     &     14.30,      292\\
A2589     &     13.89,      239&A2634     &     14.63,      210&A2644     &     12.87,      142\\
A2670     &     14.74,      391&A2715     &     14.23,      361&A2717     &     14.17,      408\\
A2721     &     14.74,      257&A2734     &     14.53,      579&A2755     &     14.82,      461\\
A2798     &     14.28,      172&A2799     &     14.07,      285&A2800     &     13.98,      304\\
A2877     &     14.69,      394&A2911     &     14.20,      200&A3093     &     13.91,      180\\
A3094     &     14.67,      450&A3111     &     12.71,       28&A3122     &     14.79,      960\\
A3126     &     14.89,      491&A3128     &     14.85,      380&A3142     &     14.72,      518\\
A3151     &     13.29,       76&A3158     &     14.97,      454&A3194     &     14.87,      373\\
A3223     &     14.51,      355&A3266     &     15.07,      480&A3334     &     14.58,      450\\
A3354     &     13.76,       94&A3360     &     14.96,      768&A3376     &     14.56,      557\\
A3381     &     13.57,      124&A3391     &     14.56,      244&A3395     &     14.76,      393\\
A3528n    &     13.99,       94&A3532     &     14.51,      198&A3556     &     14.54,      246\\
A3558     &     15.06,      240&A3559     &     14.10,      145&A3571     &     14.91,      220\\
A3574     &     14.15,      175&A3651     &     14.59,      210&A3667     &     15.07,      320\\
A3693     &     13.93,      146&A3695     &     14.59,      180&A3705     &     14.88,      267\\
A3733     &     14.46,      346&A3744     &     14.10,      192&A3809     &     14.21,      279\\
A3822     &     14.62,      132&A3825     &     14.59,      255&A3879     &     13.83,      149\\
A3880     &     14.76,      371&A3921     &     14.31,      175&A4008     &     14.02,      227\\
A4010     &     14.43,      305&A4053     &     14.31,      421&A4067     &     14.03,      139\\
AWM4      &     12.34,       20&S1157/C67 &     14.51,      498&CL2335-26 &     14.60,       79\\
DC0003-50 &     13.75,      174&Eridanus  &     13.60,      597&MKW1      &     13.22,      109\\
MKW6A     &     13.42,      169&S84       &     13.58,       70&S373      &     13.49,      138\\
S463      &     14.30,      137&S721      &     14.49,      290&S753      &     14.12,      268\\
S805      &     14.06,      184&S987      &     14.51,      151&S49-145   &     13.82,      204\\
S49-142   &     12.37,       81&N45-384   &     13.21,      144&N34-172   &     13.55,      317\\
N56-393   &     13.59,      254&N79-278   &     13.28,      149&N67-312   &     12.46,       44\\
N56-371   &     13.62,      242&N79-280   &     11.93,       10&N56-392   &     12.57,       37\\
N79-298   &     12.66,      130&N79-299B  &     13.67,      167&N67-335   &     14.06,      836\\
N79-299A  &     13.80,      247&N79-283   &     13.79,      150&N79-292   &     13.47,      814\\
\hline
\end{tabular}
 
\end{center}
\end{minipage}
\begin{multicols}{2}

\end{multicols}
%%TAB5%\vspace{-6mm}
\hspace{-13mm}
\begin{minipage}{20cm}
\renewcommand{\arraystretch}{1.2}
\renewcommand{\tabcolsep}{1.2mm}
\begin{center}%\vspace{-3mm}
TABLE 5\\%\vspace{2mm}
{\sc Continued \\}
\footnotesize 

%%%%%%%%%%%%%%%%%%%%%%%%%%%%%%%%%%%%%%%%%%%%%%%%%%%%%%%%%%%%%%%%%%%%%%%
%
%                        TABLE 5
%
%%%%%%%%%%%%%%%%%%%%%%%%%%%%%%%%%%%%%%%%%%%%%%%%%%%%%%%%%%%%%%%%%%%%%%%
\begin{tabular}{lclclc}
\hline \hline
\multicolumn{1}{c}{Name}
&\multicolumn{1}{c}{$lg\left(\frac{M}{M_{\odot}}\right),\frac{M/L_B}{(M/L_B)_{\odot}}$}
&\multicolumn{1}{c}{Name}
&\multicolumn{1}{c}{$lg\left(\frac{M}{M_{\odot}}\right),\frac{M/L_B}{(M/L_B)_{\odot}}$}
&\multicolumn{1}{c}{Name}
&\multicolumn{1}{c}{$lg\left(\frac{M}{M_{\odot}}\right),\frac{M/L_B}{(M/L_B)_{\odot}}$}
\\
\multicolumn{1}{c}{}
&\multicolumn{1}{c}{}
&\multicolumn{1}{c}{}
&\multicolumn{1}{c}{}
&\multicolumn{1}{c}{}
&\multicolumn{1}{c}{}
\\
\multicolumn{1}{c}{(1)} 
&\multicolumn{1}{c}{(2)}
&\multicolumn{1}{c}{(3)}
&\multicolumn{1}{c}{(4)} 
&\multicolumn{1}{c}{(5)}
&\multicolumn{1}{c}{(6)}
\\
\hline
N67-333   &     13.73,      229&N67-323   &     11.81,       16&N67-317   &     13.16,      302\\
N79-270   &     12.45,       75&N79-296   &     14.01,      472&N67-329   &     12.85,      100\\
N79-297   &     13.05,       73&N79-276   &     14.31,     1215&N67-336   &     13.49,      116\\
N67-325   &     13.38,      141&N67-326   &     13.44,      369&N67-309   &     13.71,      299\\
N56-394   &     13.74,      384&N56-395   &     14.16,      633&N56-381   &     13.35,      524\\
N45-381   &     12.64,       32&N45-363   &     14.07,      359&N45-389   &     14.65,      615\\
N34-171   &     13.11,      311&N34-175   &     14.18,      136&NGC 533   &     14.11,      670\\
HCG 42    &     13.12,       62&NGC 4325  &     13.49,      565&HCG 62    &     13.94,      248\\
NGC 5129  &     13.62,      228&NGC 491   &     12.00,       34&NGC 664   &     12.62,      166\\
HG3       &     12.74,       53&HG11      &     13.52,      645&HG31      &     12.59,       35\\
HG49      &     12.74,       53&HG57      &     12.53,       32&HG63      &     13.07,      149\\
HG83      &     12.37,       52&HG73      &     13.72,      263&HG94      &     12.45,       51\\
HG109     &     12.67,       59&HG120     &     12.95,      127&HG123     &     12.94,       72\\
HG138     &     11.34,        4&HG158     &     11.83,       21&HG165     &     11.80,       51\\
HG167     &     12.57,       59&HG175     &     12.59,       72&HG178     &     12.70,       97\\
HG185     &     13.23,      631&HG187     &     12.87,      125&HG200     &     12.36,      133\\
HG201     &     12.46,       98&HG212     &     13.41,      211&HG214     &     12.94,      275\\
HG223     &     11.69,       29&HG226     &     13.11,      190&HG232     &     12.98,      124\\
HG234     &     12.77,       30&HG246     &     12.85,       89&HG249     &     12.80,      291\\
HG246     &     12.94,       84&HG261     &     13.28,      188&HG309     &     13.11,      135\\
HG311     &     12.68,      109&HG322     &     11.93,       53&HG331     &     12.65,       84\\
HG333     &     11.82,       67&HG348     &     13.81,      608&HG348     &     13.64,      539\\
HG349     &     12.46,       16&HG361     &     12.38,       75&HG374     &     12.66,      102\\
HG402     &     12.36,       65&HG421     &     12.34,       30&HG429     &     11.99,       66\\
HG439     &     12.44,      283&HG454     &     12.13,       83&HG465     &     11.71,       38\\
HG473     &     14.18,      450&HG507     &     12.84,      137&HG490     &     12.12,       49\\
HG491     &     12.10,      167&HG500     &     12.54,      132&HG508     &     13.11,      165\\
HG511     &     12.85,      660&HG507     &     12.79,      111&HG525     &     12.45,      100\\
HG546     &     13.03,      828&HG545     &     13.48,      184&HG550     &     12.64,       88\\
HG557     &     13.29,      168&HG545     &     13.27,      140&HG566     &     12.87,      140\\
HG574     &     13.39,      176&HG595     &     12.49,      122&HG580     &     12.69,       80\\
HG576     &     11.73,        9&HG594     &     12.92,      167&HG595     &     12.67,      117\\
HG601     &     12.27,       24&HG602     &     13.13,      390&HG610     &     13.00,      225\\
HG574     &     13.15,      189&HG607     &     13.47,      582&HG608     &     14.09,      425\\
HG611     &     13.09,      147&HG610     &     13.21,      296&HG614     &     11.71,       97\\
HG617     &     12.20,      161&HG619     &     13.75,      753&HG622     &     14.53,      859\\
HG623     &     12.27,       57&HG626     &     12.37,       36&HG630     &     12.32,       35\\
HG636     &     13.22,      329&HG647     &     12.92,      316&HG638     &     12.41,       25\\
HG639     &     11.10,       29&HG641     &     12.54,      108&HG653     &     13.02,      189\\
HG680     &     12.15,       28&HG690     &     12.59,      156&HG694     &     12.94,      184\\
HG703     &     12.05,       25&HG704     &     13.02,      514&HG710     &     12.93,       72\\
HG712     &     11.85,       26&HG718     &     12.81,      610&HG722     &     13.29,      118\\
HG731     &     13.02,       75&HG737     &     12.81,      117&HG745     &     13.26,      278\\
HG743     &     12.44,       66&HG745     &     13.41,      301&HG758     &     11.89,       71\\
HG759     &     12.37,      119&HG806     &     12.89,      133&HG799     &     12.66,       51\\
HG806     &     13.02,      124&HG806     &     12.82,      110&HG830     &     13.30,      395\\
HG868     &     12.25,       12&HG872     &     12.49,       55&HG904     &     12.68,      151\\
HG910     &     13.74,      293&HG910     &     13.95,      306&HG913     &     13.78,      185\\
HG926     &     13.78,      315&HG939     &     13.15,      105&HG949     &     13.85,      590\\
HG949     &     13.51,      282&HG959     &     13.02,      148&HG971     &     13.38,      171\\
HG1008    &     12.60,       63&HG1017    &     12.75,      276&HG1037    &     12.68,       60\\
HG1060    &     12.31,       35&HG1057    &     13.55,      481&HG1060    &     12.29,       24\\
HG1062    &     13.35,      197&HG1071    &     13.02,      180&HG1073    &     13.40,      309\\
\hline
\end{tabular}
 
\end{center}
\end{minipage}
\begin{multicols}{2}

\section{RESULTS}

We computed the values of the mass--to--light ratio for all systems.
As for clusters and poor systems, we average COSMOS and APS
luminosities when both are available, and then we average the two
alternative luminosity estimates, $L_{B,c}$ and $L_{B,f}$, to obtain a
single value for each system. In particular, for clusters we consider
values coming from G00, too.  In Table~5 we list the values of $M$ and
$M/L$ for all 294 systems of the combined sample.

Table~6 summarizes our results, listing the median values of $M$ and
$M/L$, and 90\% c.l., for all the samples we consider.  The general
feeling is that $M/L_B$ increases with system mass (cf.  Figure~8).

\end{multicols}
%%TAB6%\vspace{-6mm}
\hspace{-13mm}
\begin{minipage}{20cm}
\renewcommand{\arraystretch}{1.2}
\renewcommand{\tabcolsep}{1.2mm}
\begin{center}%\vspace{-3mm}
TABLE 6\\%\vspace{2mm}
{\sc M/L for Different Samples \\}
\footnotesize 

%%%%%%%%%%%%%%%%%%%%%%%%%%%%%%%%%%%%%%%%%%%%%%%%%%%%%%%%%%%%%%%%%%%%%%%
%
%                        TABLE 6
%
%%%%%%%%%%%%%%%%%%%%%%%%%%%%%%%%%%%%%%%%%%%%%%%%%%%%%%%%%%%%%%%%%%%%%%%
\begin{tabular}{lrrcr}
\hline \hline
\multicolumn{1}{c}{Sample}
&\multicolumn{1}{c}{$N_S$}
&\multicolumn{1}{c}{$M$}
&\multicolumn{1}{c}{$M/L_{B_c}$, $M/L_{B_f}$}
&\multicolumn{1}{c}{$M/L_B$}
\\
\multicolumn{1}{c}{}
&\multicolumn{1}{c}{}
&\multicolumn{1}{c}{$10^{13}\,h^{-1}\,M_{\odot}$}
&\multicolumn{1}{c}{$h\,(M/L_B)_{\odot}$}
&\multicolumn{1}{c}{$h\,(M/L_B)_{\odot}$}
\\
\multicolumn{1}{c}{(1)} 
&\multicolumn{1}{c}{(2)}
&\multicolumn{1}{c}{(3)}
&\multicolumn{1}{c}{(4)}
&\multicolumn{1}{c}{(5)}
\\
\hline
A-CL&52&$20.32^{+5.89}_{-4.71}$&$262^{+43}_{-54}$, $238^{+28}_{-51}$&\\
C-CL\tablenotemark{a}&89&$23.39^{+8.85}_{-4.51}$&$276^{+31}_{-32}$, $252^{+24}_{-32}$&\\
CL&119&24.90$^{+7.23}_{-5.87}$&$270^{+32}_{-29}$, $245^{+10}_{-32}$&$246^{+31}_{-26}$\\
PS&43&$ 3.09^{+1.12}_{-1.47}$&$223^{+62}_{-80}$, $222^{+29}_{-86}$&$228^{+26}_{-84}$\\
PG &513&$0.55^{+0.15}_{-0.10}$&&$171^{+31}_{-37}$\\
PG5&208&$1.00^{+0.22}_{-0.15}$&&$197^{+38}_{-26}$\\
PG7&112&$1.67^{+0.38}_{-0.44}$&&$230^{+38}_{-37}$\\
HG &475&$0.34^{+0.10}_{-0.06}$&&$100^{+19}_{-16}$\\
HG5&190&$0.68^{+0.11}_{-0.18}$&&$123^{+15}_{-21}$\\
HG7&103&$1.02^{+0.31}_{-0.30}$&&$131^{+25}_{-22}$\\
GROUP&132&$0.63^{+0.15}_{-0.15}$&&$122^{+15}_{-20}$\\
\hline
\end{tabular}
 
\end{center}
\vspace{-2mm}
{\footnotesize\parindent=3mm 
$^a$~We report here the values obtained by G00
for clusters by using COSMOS.}

\end{minipage}
\begin{multicols}{2}

For a more quantitative analysis we avoid of fitting the behavior of
$M/L$ vs. $M$ or $L$ because $M/L$ is defined as a function of $M$ and
$L$, and therefore that would mean working with correlated quantities
(cf.  Mezzetti, Giuricin, \& Mardirossian 1982; Girardi et
al. 1996). Rather, we directly examine the $M$--$L$ relation.

%\end{multicols}
%\begin{figure}
\includegraphics{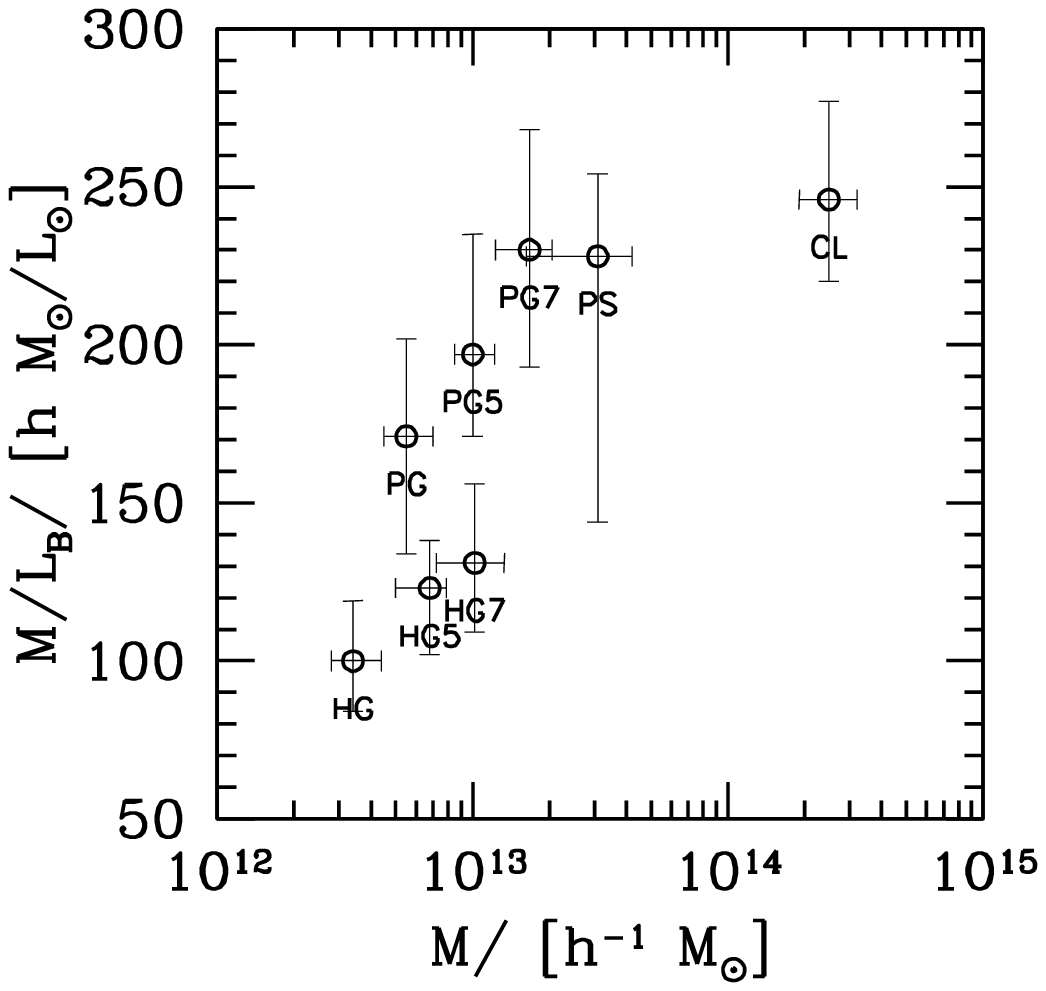}
$\ \ \ \ \ \ $\\
\vspace{8truecm}
$\ \ \ $\\
%\vspace{1cm}
{\small\parindent=3.5mm {Fig.}~8.---
Behavior of mass--to--light ratio vs. cluster
mass for the sample of clusters (CL), poor systems (PS), percolation
and hierarchical NOG groups of different richness (PG and HG,
respectively).  Circles are median values with $90\%$ c.l.  error
bars.
}
\vspace{5mm}
%\begin{multicols}{2}

%\end{multicols}
%\begin{figure}
\includegraphics{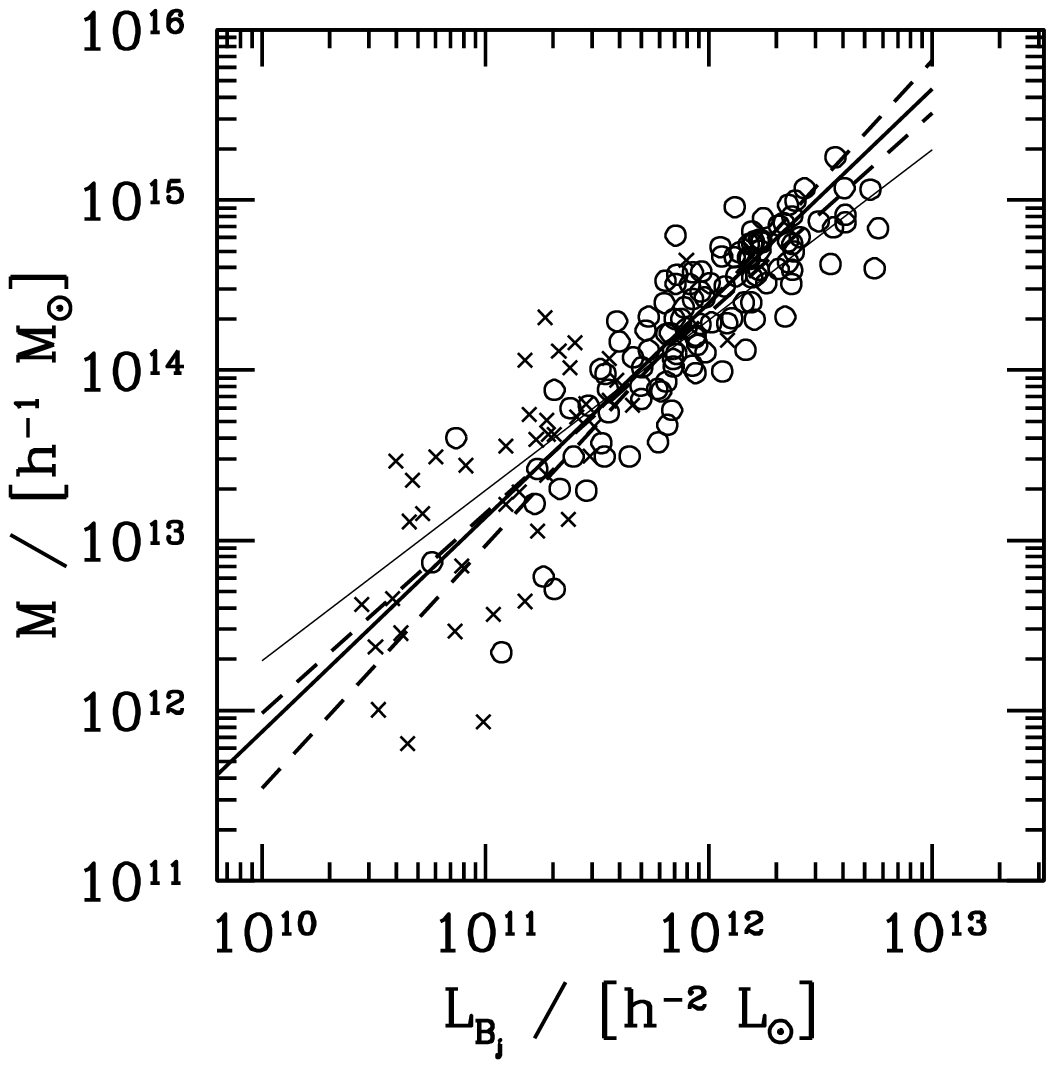}
$\ \ \ \ \ \ $\\
\vspace{8truecm}
$\ \ \ $\\
%\vspace{1cm}
{\small\parindent=3.5mm {Fig.}~9.---
Relation
between mass and luminosity for the
combined sample of clusters (CL, circles) and poor
systems (PS, crosses).
Heavy lines represent the linear fits: dashed lines give the
direct and the inverse
fits, while  the solid line gives the bisecting line.
The faint line is the $M\propto L_{B_J}$ relation.
}
\vspace{5mm}
%\begin{multicols}{2}

First, we consider together clusters and poor systems, analyzing a
combined sample (CL+PS) of 162 systems.  Figure~9 shows the
$M$vs.$L_{B_j}$ relation.  As the errors are comparable, we fit the
regression line into the logarithmic plane by using the unweighted
bisecting fit (cf. Isobe et al. 1990):
\begin{equation}
{M\over M_{\odot}}\,=\,10^c\cdot \left({L_{B_j}\over L_{B_j,\odot}}\right)^d.
\end{equation}
\noindent We obtain $c =-1.476 \pm 0.756$ and $d=1.321 \pm 0.063$.
Similar results are obtained by considering $L_{B_j,c}$ and $L_{B_j,f}$
separately, i.e.  $d=1.312\pm0.07$ and $d =1.293\pm0.056$,
respectively, both larger than one at more than the $3\sigma$ level.

Then we extend our analysis to NOG groups.  Figure~10 combines results
for all NOG groups with those for the CL+PS systems.  Both NOG
catalogs turn out to show a continuity with other systems, although
PGs seem to have larger $M/L$ ratios.

%\end{multicols}
%\begin{figure}
\includegraphics{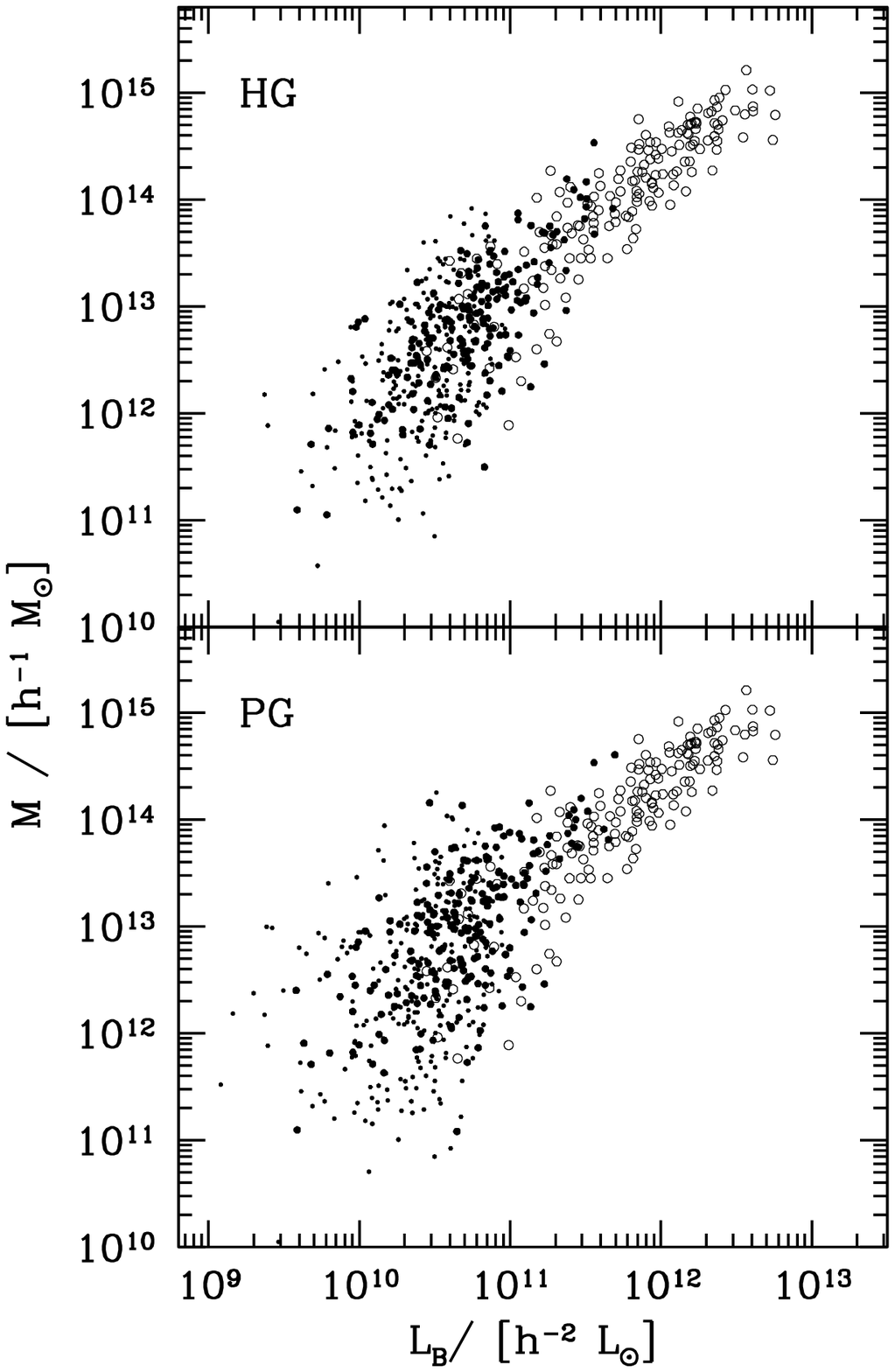}
$\ \ \ \ \ \ $\\
\vspace{15truecm}
$\ \ \ $\\
%\vspace{1cm}
{\small\parindent=3.5mm {Fig.}~10.---
Relation
between mass and luminosity for groups (solid circles, the largest
ones for groups with at least five members)
and other systems (open circles). Both HG and PG catalogs
are represented (top and bottom panels, respectively).}
\vspace{5mm}
%\begin{multicols}{2}

The analysis of our fiducial, combined sample of 294 systems
(CL+PS+GROUP), which considers only groups common to both PG and HG
catalogs with at least five members, gives:
\begin{equation}
{M\over M_{\odot}}\,=\,10^{-1.596\pm0.381}\cdot \left({L_{B}\over L_{B,\odot}}\right)^{1.338\pm0.033}.
\end{equation}
\noindent Similar results are obtained if we consider also all common
groups ($d=1.349 \pm 0.028$ for a combined sample of 458 systems), or
those with at least seven members ($d=1.309 \pm 0.036$ for a combined
sample of 231 systems).

%\end{multicols}
%\begin{figure}
\includegraphics{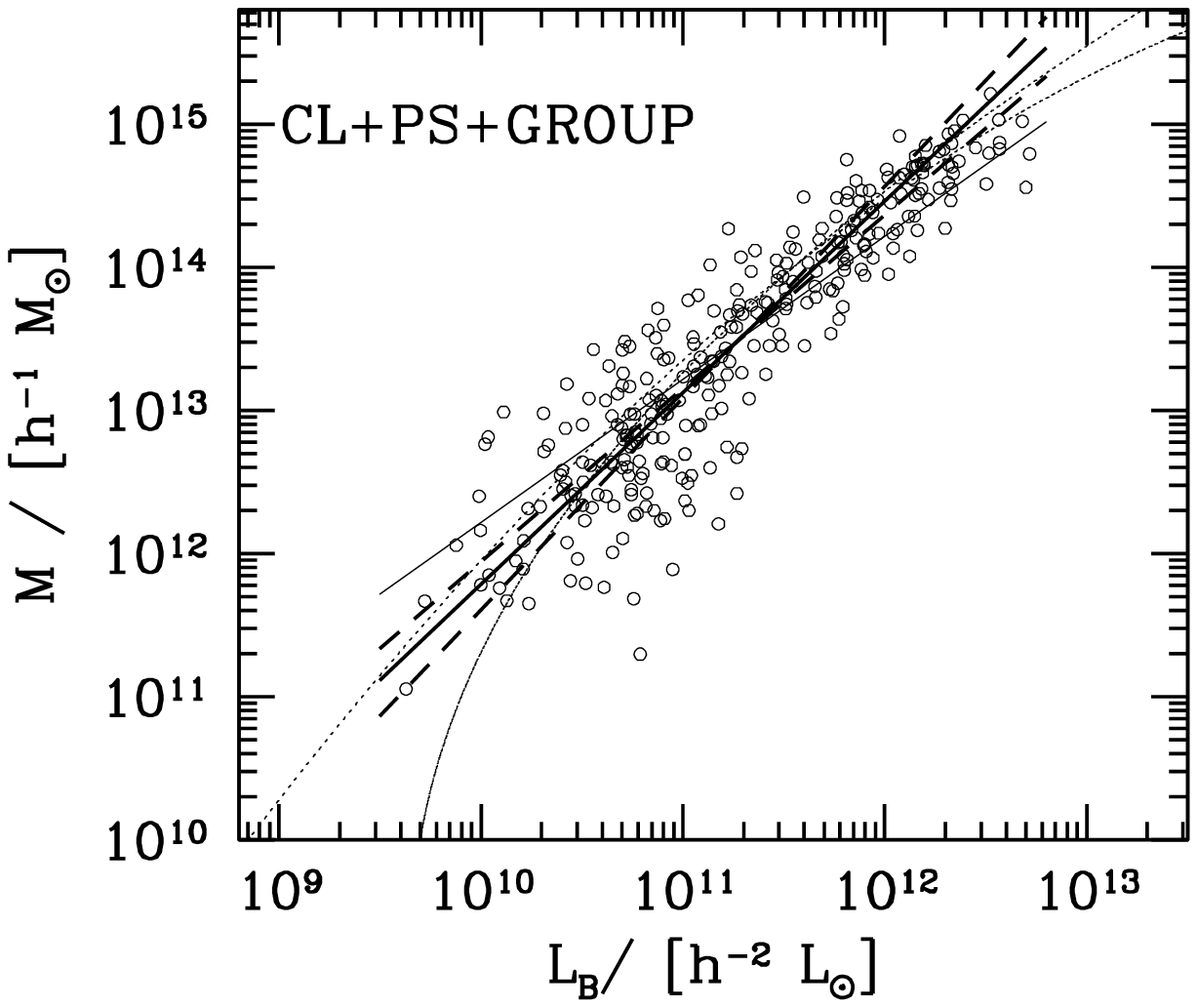}
$\ \ \ \ \ \ $\\
\vspace{8truecm}
$\ \ \ $\\
%\vspace{1cm}
{\small\parindent=3.5mm {Fig.}~11.---
Relation between mass and luminosity for our
combined sample of clusters, poor systems, and NOG groups.  Heavy
lines represent the linear fits: dashed lines give the direct and the
inverse fits, while the solid line gives the bisecting line.  The
faint solid line is the $M\propto L_{B}$ relation.  The two faint
dotted lines represent the two quadratic fits obtained by minimizing the
scatter on one or the other variable.
}
\vspace{5mm}
%\begin{multicols}{2}

Although the above straight line approach can be very useful to show
that mass increases faster than luminosity, it might not be adequate
to describe the $M$--$L$ relation in such a wide dynamical range, from
poor groups to very rich clusters. We also attempt a quadratic fit
for the two extreme situations; minimizing the scatter on $M$--axis we
obtain:
\begin{equation}
lg(M/M_{\odot})= -17.04+4.26\cdot lg(L_B/L_{B,\odot})-0.14\cdot lg(L_B/L_{B,\odot})^2,
\end{equation}
\noindent and minimizing the scatter on $L$--axis we obtain:
\begin{equation}
lg(L_B/L_{B,\odot})= 17.16-1.61\cdot lg(M/M_{\odot})+0.09\cdot lg(M/M_{\odot})^2.
\end{equation}
\noindent The first quadratic fit very closely resembles the direct
linear fit, while the second one shows a more pronounced change in the
slope of the $M$-$L$ relation; both fits show a steeper slope in the
low--mass range (cf. Figure~11).  The above results are obtained by
imposing on the dependent variable the same percent errors as masses.
Other fits with fixed errors (e.g., of $50\%$ or $80\%$) give
different numerical results but with the same qualitative behavior.
Only by having a better knowledge of error on group quantities could
we arrive at more conclusive results. In particular, although we have
not found any statistical confirmation of this, we suspect that in the
case of groups the errors on mass could be larger than the nominal
statistical ones, being due to spurious groups and/or interlopers, and
thus larger (in percent terms) than the errors on luminosity. This
possibility could explain the visual impression for a left vertical
selection boundary in the plots of $M$vs.$L$.

\section{DISCUSSION}

\subsection{Comparison with Previous Results}

Our estimate of $M/L$ (A-CL) for clusters is fully consistent with
that obtained by G00 (C-CL); we refer to G00 for other useful
discussions about clusters.  As for poor systems, recent studies give
values of $M/L_B=188$--$254$ \ml for groups with $\sigma_v=164$--$274$
\ks (Ramella, Pisani, \& Geller 1997; Tucker et al.  2000; Carlberg et
al. 2001a; Hoekstra et al. 2001a).  These results are roughly
consistent with richer NOG groups, PG5 and PG7, which have comparable
$\sigma_v$ (cf. Table~4), while the whole NOG group catalogs, which
describe the local Universe very deeply, are, as expected,
characterized by less massive systems, with smaller $M/L$.

As for the $M$--$L$ relation, analyzing 89 clusters with homogeneous
mass and luminosity estimates, G00 found that mass has a slight but
significant tendency to increase faster than the luminosity, $M\propto
L_{B_j}^{\mbox{\rm 1.2--1.3}}$, where mass and luminosity are computed
within the virial radius.  Although this result agrees with those
indirectly recovered by fundamental plane analyses (Schaeffer et
al. 1993; Adami et al. 1998a), there is a general absence of direct
evidence for a correlation between $M/L$ and cluster properties (e.g.,
Dressler 1978; David et al. 1995; Carlberg et al. 1996; Fritsh \&
Buchert 1999; but see Adami et al. 1998b).  G00 pointed out the need
for a rather large sample spanning a large dynamical range and
homogeneous analysis to detect such a small effect.

Here, with respect to G00, we consider a sample three times
larger and covering a wider dynamical range ($\sim$ two times larger
in logarithmic scale).  Our results are fully consistent with those
obtained by G00, but with a stronger statistical significance ($\sim
10 \sigma$ vs. $\sim 3\sigma$, according to face
values). Interestingly, these results have found support in some
independent recent studies.  As for groups identified in CNOC2,
Carlberg et al. (2001a) find evidence that $M/L$ increases with
increasing group velocity dispersion, and Hoekstra et al. (2001a) have
noted that the typical group $M/L$ is smaller than that of CNOC
clusters (Carlberg et al. 1996).  However, the issue is far from being
clear: e.g., Hradecky et al. (2000) have recently claimed that $M/L$
is roughly independent of system mass (but, indeed, the seven points
in their Figure~5 could also allow an increase of $M/L$).

New insights could come from a very different approach, i.e.  from
preliminary results of the correlation between the red galaxy
distribution and the dark matter distribution as measured by the
lensing signal (Hoekstra et al. 2001b; Wilson, Kaiser, \& Luppino
2001). Pioneering results support the hypothesis that red light traces
mass on scales from 0.2 \h to very large scales and it is probable
that in the future it will be possible to make a comparison with
dynamical results.

\subsection{Reliability and Caveats of Observational $M/L$}

As for the robustness of our results, several tests for luminosity
estimates were computed by G00 in their cluster analysis (cf. their
\S~6.3). In particular, they showed the small effect of changing the
analytical form and/or the parameters of the luminosity function in
the extrapolation to faint galaxies. Here, this correction is very
small for groups and poor systems, which are very close and so very
deeply sampled. 

Indeed, as for luminosity estimates, the most important correction
concerns the\\ fore/background problem, and, in fact, following G00, we
use two alternative corrections, leading to two alternative luminosity
estimates ($L_{B_j,c}$, and $L_{B_j,f}$ in \S~3.2). 

In dealing with poor, possibly spiral rich, galaxy systems, it is
worthwhile discussing the correction applied for the internal
reddening of galaxies, although this is often neglected in analyses of
the M/L ratio for galaxy systems (e.g., Hradecky et al.  2000).  In
this study, following G00, we adopt a mean correction of
$A_{B_j}=0.1$ mag for clusters: this is a compromise between the mean
correction of $A_B\sim0.3$ mag for galaxies of the Third Reference
Catalogue and the value of $A_B=0$ mag for early--type galaxies (de
Vaucouleurs et al. 1991).  As for loose groups of NOG, where the
fraction of early galaxies is comparable to that of the field
($f_e\sim 0.2$), the adopted magnitudes are already corrected for
internal absorption (Paturel et al. 1997, cf. also Bottinelli et
al. 1995).  The recent study by Tully et al. (1998) on global
extinction agrees with corrections suggested by de Vaucouleurs et
al. (1991) and Bottinelli et al. (1995): they find negligible
extinction in lenticulars and $A_B\sim 1.8$ in highly inclined spirals
(cf. with $A_B\sim 1.5$--$1.67$ by de Vaucouleurs et al. 1991 and
Bottinelli et al. 1995).  Indeed, some specific studies of highly
inclined spiral galaxies could suggest higher extinction values,
finding $A_B=2$--3 mag in the center and then a rapid drop with radius
(cf. Jansen et al. 1994, see also Kuchinski et al. 1998).  Even
supposing that we are underestimating the internal extinction of
spirals by a factor of two, the group $M/L$ is presently overestimated
at most by a factor of $\sim 30\%$ and the slope of the $M$--$L$
relation is presently slightly underestimated (e.g., we would obtain
$M \propto L_B^{1.4}$ by applying a $30\%$ correction to NOG group
luminosities).

Finally, as for the robustness of our luminosity estimates, we go well
beyond G00 results on one particular point.  In fact, while G00
results are strongly based on the COSMOS catalog, we have shown that
luminosities coming from two different catalogs (COSMOS and APS) are
really comparable, suggesting that no systematic effect, connected to
a particular catalog, pollutes our results.

The most important systematic uncertainty concerns mass estimates,
since our application of the virial theorem assumes that, within each
system, mass distribution follows galaxy distribution.  For clusters
this assumption is supported by several independent analyses using
optical and X--ray data, as well as gravitational lensing phenomena
(e.g., Durret et al. 1994; Narayan \& Bartelmann 1996; Carlberg, Yee,
\& Ellingson 1997; Cirimele, Nesci, \& Trevese 1997), but we must
recognize that the issue is far from being clear for poor systems. The
absence of luminosity segregation of galaxies in the velocity space
(Giuricin et al. 1982; Pisani et al. 1992) suggests that the effect of
dynamical friction in slowing down galaxies with respect to dark
matter is very poor. However, analyzing CNOC2 groups, Carlberg et
al. (2001a) have recently shown that light might be much more
concentrated than mass. If that also galaxy number distribution is
more concentrated than mass, our virial mass estimates for very poor
systems would be underestimated (Merritt 1987) leading to a steeper
$M$--$L$ relation with respect to the true one. Therefore, we stress
that our results are strictly correct only if galaxy distribution
traces mass within each individual galaxy system.

In the specific case of groups, one could suspect that the particular
selection procedure used biases $M/L$; i.e.  groups having at least
three luminous galaxies (with $B<14$ mag) could be on average more
luminous and so with systematically smaller $M/L$.  In order to check
for this possible bias, we consider several subsamples of groups
having a progressively higher luminosity for the third--brightest
galaxy: i.e.  groups with at least 3 galaxies more luminous than 13.5
mag (13.0 mag, 12.5 mag, 12.0 mag).  For percolation groups, having
median $M/L_B=171$\mll, we obtain for the subsamples, respectively:
$M/L_B=173$, 174, 158\mll; and for hierarchical groups, having median
$M/L_B=100$\mll, we obtain: $M/L_B=109$, 104, 119 \mll.  Since there
is not a clear trend for a systematic decrease of $M/L$, we expect
that this bias, if any, is negligible.

Unfortunately, our results are not connected in any obvious way to the
relation between stellar and total mass in galaxy systems.  Indeed, it
would be more appropriate to make a direct analysis based on infrared
light, which seems a better tracer of stellar mass (e.g., Gavazzi,
Pierini, \& Boselli 1996). Here we only attempt to infer some
conclusions, after discussing the morphological content of different
systems.

As for clusters, it is well known that the fraction of late--type
galaxies decreases with the local density (Dressler et 1980) and
increases with the distance from the cluster center (Whitmore,
Gilmore, \& Jones 1993). In fact, clusters are characterized by the
presence of color gradients in the radial direction (Abraham et
al. 1996; cf. also Fairley et al. 2001). When working within a
physical radius, rather than in a fixed spatial radius, the morphology
content seems to be roughly independent of mass (e.g., Whitmore et al.
1993; Fairley et al. 2001).  In this context, our analysis performed
within the virial radius suggests that we are dealing with comparable
galaxy populations, so that our observed $M$--$L$ relation should also
be valid in other magnitude bands.

The situation could be different for poor systems, where the question
is far from being clarified. For instance, it is not clear if systems
with very low X--ray luminosity are characterized by a very small
fraction of bulge--dominated and/or red galaxies; cf. Balogh et
al. 2001 and Fairley et al. 2001 for different results.  As for loose
groups, Carlberg et al. (2001b) have shown the presence of color
gradients in massive galaxy groups and Tran et al. (2001) have found
evidence for a morphology--radius relation in X--ray detected groups,
similar to that of clusters. Instead, no trace of color gradients has
been found for less massive groups (Carlberg et al. 2001b).  In our
case, the fraction of early galaxies in NOG groups is comparable with
that of the field ($f_e\sim 0.2$, to be compared to $f_e\sim
0.5$--$0.75$ in clusters, e.g. Oemler 1974).  This could mean that NOG
groups are different from clusters in their morphological content or
that they are similar, but we are looking at the the combined effect
of color gradients and a sampling area that is very large with respect
to clusters ($\sim 2 \times R_{vir}$). Whatever the reality is, one
needs to apply a correction to pass from blue to infrared
luminosities.

Assuming $(B-H)\sim 3.75$ and $\sim 3.0$ for early type and late type
galaxies, respectively (Fioc \& Rocca-Volmerange 1999), as well as
$(B-H)_{\odot}=2.1$ (Wamsteker 1981), we find that $L_H^e=4.6 L_B^e$
for luminosity of early--type galaxies and $L_H^l=2.3 L_B^l$ for
luminosity of late type galaxies.  Then, when assuming that typical
blue galaxy luminosities in clusters are roughly comparable for early
and late type galaxies (e.g. Sandage, Binggeli, \& Tammann 1985;
Andreon 1998) and that the early galaxy fraction goes from 0.75 to 0.2
for spiral--poor clusters and groups, respectively, we obtain $L_H\sim
4.0 L_B$ and $L_H\sim 2.8 L_B$ for clusters and groups, respectively.
Therefore, we expect that $M/L_H$ for groups (GROUP sample) will still
be lower than for clusters (CL sample, cf. Table 6) by $\sim 40\%$,
compared with a factor two difference in $M/L_B$.

\subsection{Comparison with Theoretical Results}

The assumption that $M/L$ within galaxy clusters is typical of the
Universe as a whole leads to an estimate of the matter density
parameter $\Omega_m$, i.e. $\Omega_m=(M/L)\cdot \rho_L/\rho_c$, where
$\rho_c$ is the critical density, and $\rho_L$ is the typical
luminosity density of the Universe, as generally determined on field
galaxies (Oort's method, e.g. Bahcall et al. 1995; Carlberg et
al. 1996; G00).  However, both assumptions that luminosity is
conserved when field galaxies fall into a cluster and that galaxy
formation is the same in all environments are questionable.  Recently,
the combination of cosmological numerical simulations and semianalytic
modeling of galaxy formation have faced the question of
galaxy--systems $M/L$ in a more realistic way (e.g., Kauffmann et
al. 1999; Bahcall et al. 2000; Benson et al. 2000; Somerville et
al. 2001).  Here we attempt a comparison with the theoretical results.

Figure~12 compares our observational results with the theoretical
predictions of Kauffmann et al. (1999) and Benson et al. (2000), who
both recovered the behavior of $M/L_B$ vs. halo mass of galaxy
systems in the framework of cold dark matter (CDM) models for two
alternative cosmologies: a low--density model with $\Omega_m=0.3$ and
$\Omega _{\Lambda }=0.7$ ($\Lambda CDM$), and a high--density model with
$\Omega_m=1$ and shape parameter $\Gamma=0.21$ ($\tau$CDM).

%\end{multicols}
%\begin{figure}
\includegraphics{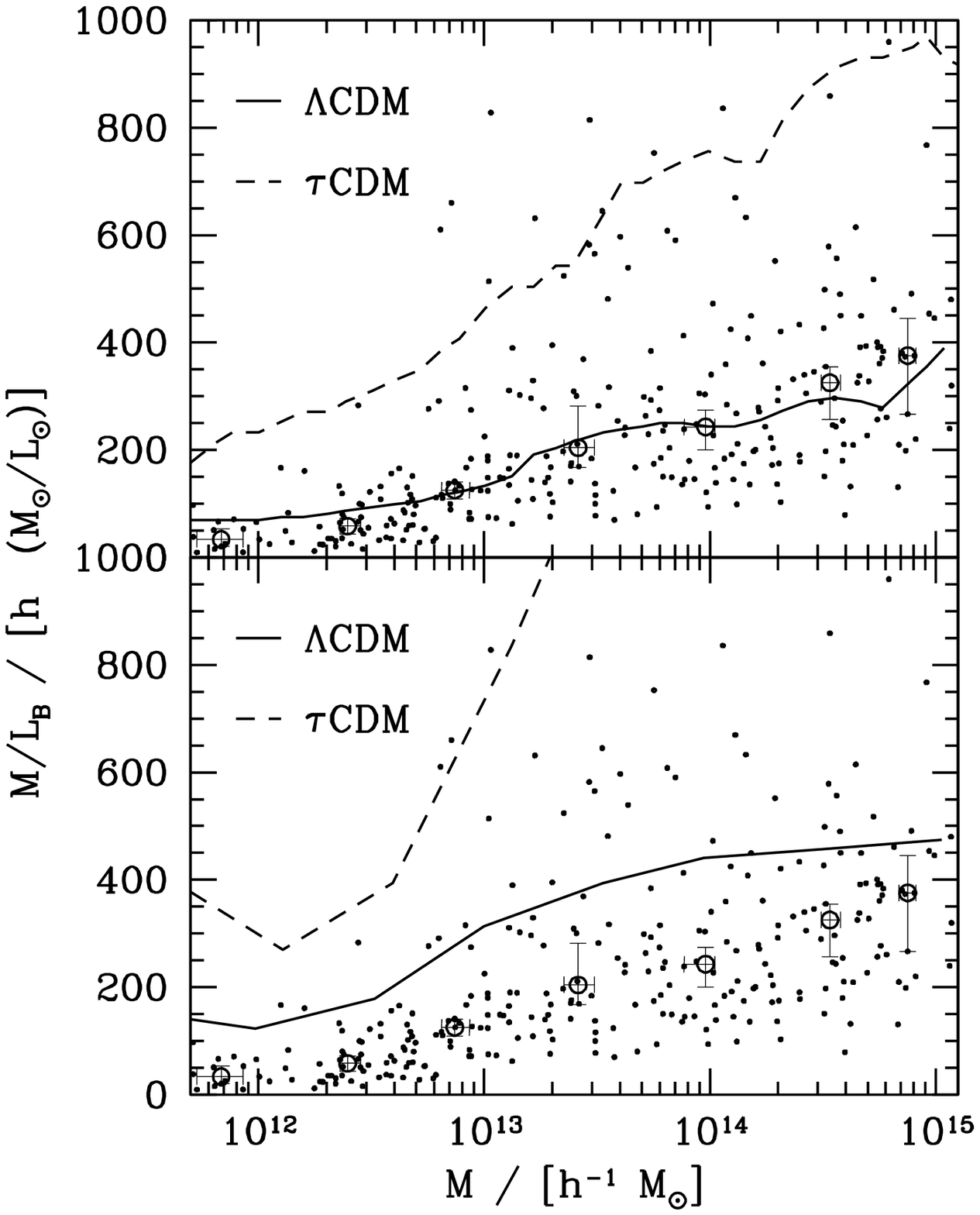}
$\ \ \ \ \ \ $\\
\vspace{13truecm}
$\ \ \ $\\
%\vspace{1cm}
{\small\parindent=3.5mm {Fig.}~12.---
Comparison between the observational behavior of
mass--to--light of galaxy systems and the theoretical predictions of
Kauffmann et al. (1999, top panel) and Benson et al. (2000, bottom
panel); see text.  When plotting Kauffmann et al.'s results we assume a
closure value for the Universe $M/L_B[Universe]=1350$ \ml (from the
luminosity density $\rho_l\sim2\times10^8~hL_{\odot}Mpc^{-3}$ by
Efstathiou, Ellis, \& Peterson 1988). For comparison, Benson et
al. quoted a mean value of $M/L_B=1440$ \ml in their simulation as a
whole in the $\tau CDM$ cosmology.  Points represent individual data
for our combined sample, while circles show median values with $90\%$
c.l. error bands.
}
\vspace{5mm}
%\begin{multicols}{2}

The $\Omega_m=1$ cosmology is reported here for the sake of
completeness. Also without taking into account the small--mass range, where
we should recompute the mass of NOG groups for this kind of model,
$\Omega_m=1$ cosmology is clearly rejected according to the 
our observational results.

However, the preferred value of $\Omega_m$ is not obvious, since the
model with $\Omega_m=0.3$ by Kauffmann et al.  fits the data well,
while the results of Benson et al. suggest a smaller value.  Indeed,
the predicted value of $M/L$ shows large differences among the
predictions of different authors and can vary considerably in many
theoretical details, e.g. the correction of dust extinction
(cf. Somerville et al. 2001).  Besides normalization, also the whole
behavior of $M/L$ can be very useful in constraining theoretical
results: e.g., the results obtained by Kauffmann et al. reproduce the
steepness of the observational increase of $M/L$ with halo mass, while
the results by Benson et al. show a flatter behavior.

Finally, we discuss the very recent results by Marinoni and Hudson
(2001), who derive the behavior of $M/L$ by using the analytical
approach of Press \& Schechter (1974) and the observational luminosity
function for galaxy systems: their prediction for the $\Lambda CDM$
model agrees well with our findings in the $10^{13}$--$10^{14}$ \msun
range, but they obtain a steeper slope in the high--mass range, and an
slope inversion in the low--mass range.

\section{Summary and Conclusions}
We analyze the mass--to--light ratios of galaxy systems from poor
groups to rich clusters by considering virial mass estimates and blue
band luminosities.  

We extend the previous work of G00, where they computed $B_j$ band
luminosities derived from the COSMOS catalog (Yentis et al. 1992) for
a sample of 89 galaxy clusters, with virial mass homogeneously
estimated by G98.

In this study we consider another 52 clusters having virial masses
estimated by G98, a sample of 36 poor clusters proposed by L96, and a
sample of 7 rich groups well analyzed by ZM98.  For each poor system
we select member galaxies and compute virial mass as performed by
G98. For all systems we compute $B_j$--band luminosity by using both
the APS catalog (Pennington et al. 1993) and the COSMOS catalog with
the same procedure as that adopted by G00.  Both mass and luminosity
for each object are computed within the virial radius, in order to
consider comparable physical regions for systems of different
mass. The advantage of this procedure lies in the fact that one can
compare regions with similar dynamical status and galaxy
populations. By also taking into account the results of G00, we obtain
a sample of 162 galaxy systems having homogeneous mass and luminosity
estimates.

To extend our data base, we consider the two group catalogs identified
in the NOG catalog by Giuricin et al. (2000), based on two different
group identification algorithms, a percolation one and a hierarchical
one ($\sim 500$ groups for each catalog).  We compute mass and blue
band luminosity for each group, homogeneizing our results to those of
other systems as much as possible; in particular, we rescale mass and
luminosity to the central, possibly virialized, group region.

To avoid possible spurious groups, we consider the subsample of 132
NOG groups identified in both catalogs and having at least five
members. We combine these groups with clusters and poor systems to
obtain a fiducial combined sample of 294 systems spanning a very large
dynamical range ($\sim 10^{12}$--$10^{15}$ \msun).

We find that mass increases faster than luminosity.  By using the
bisecting unweighted procedure, the analysis of the combined sample
gives:
\begin{equation}
{M\over M_{\odot}}\,=\,10^{-1.596\pm0.381}\cdot \left({L_{B}\over
L_{B,\odot}}\right)^{1.338\pm0.033}.
\end{equation}
\noindent Consistent results are recovered by using the more
homogeneous subsample, which contains only 162 clusters and poor
systems.  This result agrees with that reported by G98, confirming
the effect at a higher statistical significance (there the effect was
detected at the $\sim 3\sigma$ level).

When analyzing the combined sample with a quadratic fitting relation,
we find a tendency for a steeper slope in the low--mass range.

Finally, we compare our observational results with the theoretical
predictions with the behavior of $M/L_B$ vs. halo mass, in particular
to the behavior recently predicted by the combination of cosmological
numerical simulations and semianalytic modeling of galaxy
formation. We find a very good agreement with the result by Kauffmann
et al. (1999) for a CDM model with $\Omega_m=0.3$ and
$\Omega_{\Lambda}=0.7$.  This demonstrates that the study of the
mass-to-light ratio scaling for galaxy systems represents a useful
tool for constraining models of galaxy formation.

\acknowledgments We wish to dedicate this paper to the memory of our
friend and colleague Giuliano Giuricin who suddenly died during the
preparation of this paper.  We thank Andrea Biviano and Stefano
Borgani for useful discussions. We thank an anonymous referee for
useful suggestions.  This research has made use of the APS Catalog of
POSSI which is supported by the National Aeronautics and Space
Administration and the University of Minnesota (the database can be
accessed at $http://aps.umn.edu/$); the ROE/NRL UKST Southern Sky
Catalogue, work funded in part by the Office of Naval Research and by
a grant from the NASA ADP program (the database can be accessed at\\
$http://xip.nrl.navy.mil/www\_rsearch/RS\_form.html$); the COSMOS/UKST
Southern Sky Catalogue supplied by the Anglo--Australian Observatory;
and the NASA/IPAC extragalactic Database (NED), which is operated by
the Jet Propulsion Laboratory, California Institute of Technology,
under contract to the National Aeronautics and Space Administration.
This work has been partially supported by the Italian Ministry of
Education, University, and Research (MIUR), and by the Italian Space
Agency (ASI).

%%%%%%%%%%%% per formato preprint
\end{multicols}
%%%%%%%%%%%% per formato preprint
%\begin{multicols}{2}

\end{document}